\documentclass[authoryear,final,10pt,5p,times,epsfig]{elsarticle}

 \usepackage{epsfig}

\usepackage{amssymb}

\journal{Icarus}

\begin{document}

\begin{frontmatter}



\title{The Maximal Runaway Temperature of Earth-like Planets  }


\author{Nir J.\ Shaviv}

\address{Racah Inst.\ of Physics, Hebrew University of Jerusalem, Giv'at Ram, Jerusalem 91904, Israel}

\author{Giora Shaviv}

\address{Dept.\ of Physics, Israel Institute of Technology, Haifa 32000, Israel}

\author{Rainer Wehrse}

\address{The Institute for Theoretical Astrophysics, Heidelberg University, Germany}

\begin{abstract}

In Simpson's (1927) classical derivation of the temperature of the Earth in the semi-gray model, the surface temperature diverges as the fourth root of the thermal radiation's optical depth. No resolution to this apparent paradox was yet obtained under the strict {\em semi-gray} approximation. Using this approximation and a simplified approach, we study the saturation of the runaway greenhouse effect. 

{First we generalize the problem of the semi-gray model to cases in which a non-negligible fraction of the stellar radiation falls on the long-wavelength range, and/or that the planetary long-wavelength emission penetrates into the transparent short wavelength domain of the absorption. } 

Second, applying the most general assumptions and independently of any particular properties of an absorber, we show that the greenhouse effect saturates and that any Earth-like planet has a maximal temperature which depends on the type of and distance to its main-sequence star, its albedo and the primary atmospheric components which determine the cutoff frequency below which the atmosphere is optically thick. For example, a hypothetical convection-less planet similar to Venus, that is optically thin in the visible, could have at most  a surface temperature of 1200-1300K irrespective of the nature of the greenhouse gas.

We show that two primary mechanisms are responsible for the saturation of the runaway greenhouse effect, depending on the value of $\lambda_\mathrm{cut}$, the wavelength above which the atmosphere becomes optically thick. Unless $\lambda_\mathrm{cut}$ is small and resides in the optical region, saturation is achieved by radiating the thermal flux of the planet through the short wavelength  tail of the thermal distribution. This has an interesting observational implication, the radiation from such a planet should be skewed towards the NIR. Otherwise, saturation takes place by radiating through windows in the FIR.

\end{abstract}


\end{frontmatter}


\def\e{{\mathrm{e}}}
\def\todo#1{#1}
\newcommand {\xx} {{\bf x}}
\newcommand {\yy} {{\bf y}}
\newcommand {\nn} {{\bf n}}
\newcommand {\EE} {{\bf E}}
\newcommand {\GG} {{\bf G}}
\newcommand {\II} {{\bf I}}
\newcommand {\pp} {{\rm I\hskip -0.2em P}}
\newcommand {\block} {\vrule width 0.25 true cm height 6 pt depth 0 pt \ }
\newcommand {\IB} {{\cal I}}
\newcommand {\SB} {{\cal S}}
\newcommand {\JJ} {{\cal J}}
\newcommand {\kc} {{\it k}}
\newcommand {\kpc} {{\it k^\prime}}
\newcommand {\MG} {{ M}}
\newcommand {\MM} {{\cal M}}
\newcommand {\taud}  {\tau_{\scriptscriptstyle 3}}
\newcommand {\bra} [1] {<\hspace{-2 pt} #1 \hspace{-3 pt} \mid}
\newcommand {\ket} [1] {\mid \hspace{-2 pt} #1 \hspace{-2 pt} >}
\newcommand {\Psib} {{ \Psi}}
\newcommand {\RR} {I\hskip -4pt R}
\newcommand {\CC} {I\hskip -8pt C}
\newcommand {\sgreekbf}[1]{\ensuremath{\mbox{\boldmath$#1$}}}
\renewcommand {\SS} {{\bf S}}

\renewcommand {\Re} {{\it Re}}

\newcommand {\eff} {{\mathrm{eff}}}
\newcommand {\eq} {{\mathrm{eq}}}
\newcommand {\rad} {{\mathrm{rad}}}
\newcommand {\FIR} {{\mathrm{FIR}}}
\newcommand {\SW} {{\mathrm{SW}}}

\newcommand {\deff} {d_{\mathrm{eff}}}

\newcommand{\vdag}{(v)^\dagger}

\newcommand{\figstyle}{  }
\newcommand{\oa}{}

\newcommand{\lpeaks}{\lambda_\mathrm{peak}^{\star}}
\newcommand{\lpeakp}{\lambda_\mathrm{peak}^{\circ}}
\newcommand{\lrad}{\lambda_\mathrm{rad}}
\newcommand{\lcut}{\lambda_\mathrm{cut}}
\newcommand{\ncut}{\nu_\mathrm{cut}}
\newcommand{\tauSW}{\tau_\mathrm{SW}}
\newcommand{\tauFIR}{\tau_\mathrm{FIR}}
\newcommand{\Tsur}{T_\mathrm{sur}}
\newcommand{\kappaSW}{\kappa_\mathrm{SW}}
\newcommand{\kappaFIR}{\kappa_\mathrm{FIR}}
\renewcommand{\em}{\it}

\newcommand{\planss}{Planet.\ Sp.\ Sc.}
\newcommand{\jgr}{J.\ Geophys.\ Res.}
\newcommand{\aap}{Astron.\ Astrophys.}
\newcommand{\apj}{Astrophys.\ J.}


\section{Introduction}

The runaway of the greenhouse effect was first described by \cite{Simpson,Simpson2}. He pointed out to what appeared to be a paradox.  In his solution, the atmospheric temperature diverges as the fourth root of the optical depth of any greenhouse gas, such as water vapor. Without any stabilizing negative feedback, this divergence became to be known as the ``Simpson's paradox" (cf.\ \citealt{Goody}).

\cite{Brunt} elaborated by combining the effect of heat transfer by radiation and turbulence through the assumption that both processes can be described with additive diffusion coefficients.

Later, \cite{Sagan60} and \cite{Gold} noted that if a planet is too close to the sun, enough water vapor would evaporate to give rise to a positive feedback and heat the surface even more. This runaway effect would eventually evaporate the whole body of water and increase the atmospheric optical depth in the IR. 

\cite{King}  treated the semi-gray atmosphere while implicitly assuming  that the frequency separating the incoming shortwave (SW) and the outgoing far infrared (FIR) radiation is the same as the frequency below which the atmosphere is opaque. In addition, King  assumed that $\tauSW$, the optical depth in the visible range,  is large as well. King demonstrated that the surface temperature diverges and depends through the Chandrasekhar $H(\eta)$-function \citep{Chandra} on $\eta=\tauFIR / \tauSW$---the ratio between the Far IR optical depth and the shortwave one.

\begin{sloppypar}
Later, \cite{Shultis} extended the results of \cite{King} and \cite{Wildt}, to include anisotropic scattering.  \cite{Stibbs} extended King's results and obtained the emergent atmospheric radiation ``for any combination of values of the parameters". However, a  tacit assumption in all the above treatments of the problem is that the wavelength where the stellar and planetary radiations fields are equal, and the wavelength at which the strong IR absorption starts, are identical.
\end{sloppypar}

Stibbs defined $n=1/\eta$ as {\it ``a measure of the effectiveness with which thermal photons generated by the absorption and degrading of the dilute incident visual radiation"}.  In the case $n\ll1 $, incident photons can penetrate the atmosphere much more easily than the generated thermal photons can escape. Consequently, the thermal photons tend to congregate where they are generated, except those near the surface, and thereby establish a source function in the thermal radiation which increases with optical depth in the atmosphere to produce the greenhouse effect. Interestingly, Stibbs also found the conditions for limb brightening (when heating at the top of the atmosphere dominates the structure of the atmosphere), and demonstrated that the conditions depend solely on $\eta$.

{\oa
\cite{Komabayashi} was worried from the ``Possibility of exponential boiling-off of the ocean on planets with positive feedback leading to endless warming between the vapor pressure and the greenhouse effect" and attempted to provide a ``qualitative analysis using a model of one component-two phase system". Kombayashi was concerned with whether it is possible that there is no equilibrium solution to the problem of radiative transfer with water vapor leading to a runaway. Komabayashi  solved the problem by adopting the Eddington approximation in the grey approximation, and found a solution in which the water vapor did not prevent temperature  divergence.

}{}

  \cite{Ingersoll} discussed the possible runaway on Venus and considered only  the IR flux from below. He used the standard Simpson solution, namely: $B^{*}(\tau_{tot})={F / 2\pi}\left(2+({3 / 2})\tau_{tot}\right),$ where $F$ is the value of the upward radiative flux,  $\tau_{tot}$ is the optical depth from infinity to the surface, and $B(T)=\sigma T^4/\pi$. Ingersol   discussed the radiative-convective equilibrium model coupled to a model in which the IR absorption is due to vapors in equilibrium with its liquid or solid phase. This analysis led to the conclusion that equilibrium is impossible when the solar constant exceeds a critical value as an equilibrium model requires that the condensed phase evaporates into the atmosphere. Ingersoll concluded that a runaway is possible for  a planet with oceans at the orbit of Venus. 

\begin{sloppypar}
 The works by \cite{Simpson}, \cite{Komabayashi} and \cite{Ingersoll} gave rise to the so called SKI limit, which is the largest outgoing long-wave radiative  flux which a planet can emit while sustaining the liquid (or solid) phase at the surface. When the absorbed stellar radiation surpasses this limt, there is no equilibrium and a complete evaporation of the condensed phase occurs.
\end{sloppypar}

\cite{Rasool} discussed the accumulation of ${\rm CO}_2$ in Venus' atmosphere and its runaway. The authors applied the Eddington approximation for Planck mean of $\kappa$. Thus, the absorption was assumed to be  constant over all wavelengths,  and the  radiative-convective model for the mean absorption was implemented for all wavelengths.

\cite{Kasting88}  discussed questions regarding runaway, moist greenhouse atmospheres and the evolution of the Earth and Venus and argued that the reason for the Venus's atmosphere runaway is the higher (a factor of $1.91S_{\oplus}$) solar insolation, where $S_{\oplus}$ is the solar insolation at the mean distance of the Earth from the Sun.  Using the radiative-convective model,  he found that the runaway limit is an insolation of $1.4S_{\oplus}$. Thus, Venus accordingly, was in the past below the runaway limit having liquid water, but as the solar intensity increased, the limit was reached and induced a runaway.

The increase of  stratospheric water vapor   with surface temperature sets an even more stringent limit of $1.1S_{\oplus}$. Thus, the liquid water should have evaporated at  an earlier time. Mars on the other hand, is within the ``maximum greenhouse" limit \citep{Kasting93},  and remains outside ``the first ${\rm CO}_2$ condensation" limit.

\begin{sloppypar}
When discussing how the runway takes place, \cite{Kasting88} raised the idea that the planetary temperature cannot grow indefinitely because there always exists a sufficiently high temperature for which the thermal radiations' short wavelength tail would eventually be large enough to irradiate the stellar short wave radiation reaching the surface. This is one of the two primary ideas we elaborate upon here. 
\end{sloppypar}

{\oa 
\cite{Renno94} included the hydrological cycle in a 1D radiative-convective model. The radiation flux was calculated using a parametrization scheme and was not solved explicitly. \cite{Renno97} studied the effect of water vapors and showed that there can be two linearly stable solutions to the radiative-convective equilibrium when the net forcing is larger than a critical value. However, a finite amplitude instability can lead to a runaway greenhouse regime when the solar forcing is larger than a second critical value. The first equilibrium corresponds to an optically thin atmosphere while the second equilibrium corresponds to an optically thick, highly nonlinear atmosphere. Unfortunately, the calculation is highly uncertain because clouds were not included in the modeling. 

\cite{Nakajima} performed a study of the runaway greenhouse effect with a one-dimensional radiative-convective equilibrium model. The publication also supports the point of view that ``there is no saturation of the greenhouse effect resulting from an unlimited accumulation of greenhouse gases in the atmosphere of a planet." Note that these authors refer to the state in which oceans cannot exits in an equilibrium, as the runaway greenhouse state. This is not exactly what we discuss here. 
}{}

\begin{sloppypar}
{\oa 
\cite{Pujol} elaborated the SKI analysis by alleviating the implicit assumption of \cite{Simpson}, \cite{Komabayashi} and \cite{Ingersoll} that the absorbing greenhouse gas does so equally at all infrared wavelengths (i.e., gray atmosphere)}. The authors found several equilibria and that the anti-greenhouse effect can remove the radiation limit.  
\end{sloppypar}

The origin of the divergency in the semi-gray model is the assumption that  radiation can only escape through the optically thick IR. If so, an increase in  the optical depth requires a higher temperature gradient to carry the same heat flux from the surface.  The temperature increase would stop only if heat can be carried out by other means. For example, \cite{Weaver} extended the Goody and Yung's analysis of the Simpson's paradox by mainly considering  windows in the absorption, that is, they deviated from the original semi-gray model. {\oa We discuss this possibility and show that it does not solve the fundamental paradox. }{} Another important aspect is that a high temperature gradient can excite very efficient convection which will carry the  excess heat from the surface and limit the temperature gradient to the adiabatic one. We defer the discussion of convection to a forthcoming publication.

\begin{sloppypar}
Leaving the gray approximation, \cite{Kasting} used a band model in the diffusion approximation and obtained extremely high temperature for Venus, of the order of 1400K. 
The diffusion approximation for the radiation field in gray atmosphere was extended by \cite{Minin} to cases with heat sources in the atmosphere. This was carried out by assuming two $B(\tau)$ functions, each describing another radiative flux. Minin wrote the general equation, though he did not refer to King's results. 
\end{sloppypar}

Later, \cite{Lorenz} discussed the climate stability of Titan. Their solution for the radiative transfer included the blocking of shortwave radiation and the emission through windows in the FIR. They employed the analytical model of \cite{Samuelson} and the more advanced models developed by \cite{McKay1989}. In principle, the two treatments assumed the diffusion approximation for the thermal radiation and yielded an expression of the form $ \sigma \Tsur^4=a\left(q+{3\tau/ 4} \right)+b $,
where $T_{sur}$ is the surface air temperature, while $a,b$ and $q$ are constants which depend on Titan's parameters. $\tau$ is defined as the thermal optical depth of the atmosphere.   The above result is written so as to expose the dependence on the optical depth, showing the resemblance to Simpson's result. The same solution for the temperature was also applied by \cite{McKay}, who discussed the anti-greenhouse effect on Titan and the early Earth, an effect which arises from  a large shortwave optical depth. This was also discussed by \cite{Pujol}.

\cite{Miskolczi} repeated the calculation of the radiative transfer in the diffusion approximation, essentially rederiving the results of both Simpson and \cite{Goody} for  $\Tsur (\tau_{tot})$. Miskolczi was bothered by the non-physicality of the assumption $\tau_\mathrm{tot}\rightarrow \infty$, while relying on \cite{Milne} in claiming that the behavior is not physical. However, his revised solution still contains the problem that the temperature increased  indefinitely as the optical depth increases monotonically. 

More recently, \cite{Rutilya} succeeded to obtain analytical results for the semi-gray atmospheres while assuming the following particular assumptions that: (a) The radiation field in the spectral region of low absorption coefficient is pure solar.  (b) The emission of the atmosphere in the short wavelength range can be neglected. (c) The thermal emission of the surface is negligible in the short wavelength range.  We  show that these assumptions become invalid as the total absorption in the long wavelengths increases.

In the present paper,  we do not discuss the effect of minimal insolation leading to a runaway. Instead, we study the saturation as a function of the total optical depth.  To  this goal, we present a wavelength-dependent analytical and a full numerical solution of the semi-gray model as well as simplified analytical estimates. We start with several definitions and an exposition of the model, and then continue with the numerical results. These results demonstrate that there is a marked difference between a wavelength dependent model and either the analytical diffusion or two streams description within a two or few band model. In particular, we show that contrary to Simpson's result, the surface temperature does not increase indefinitely as the total optical depth increases to infinity, but instead  it saturates even in the classical semi-gray approximation. We explain the saturation and derive a simple approximation for the saturation temperature as a function of the model parameters.  The advantage of the semi-gray model is in its simplicity. As the  model ignores the details of the absorbers, the obtained 	 results  are quite general and do not depend on the properties of any particular molecule. 	


We begin in \S\ref{sec:model} by developing the simplest semi-gray model which can describe the saturation of the greenhouse effect. In \S\ref{sec:simpson} we discuss the conditions in which the Simpson solution is obtained. We then show in \S\ref{sec:IRwindow} how the solution is modified when a window in the FIR is introduced. We
 then describe the numerical results to the full semi-gray model in \S\ref{sec:numerical}. The anti-greenhouse effect is discussed in \S\ref{sec:anti-green} and the implications of the results are discussed in \S\ref{sec:discussion}. 

\section{The Semi-Gray model}
\label{sec:model}

\begin{figure}
\center{\epsfig{file=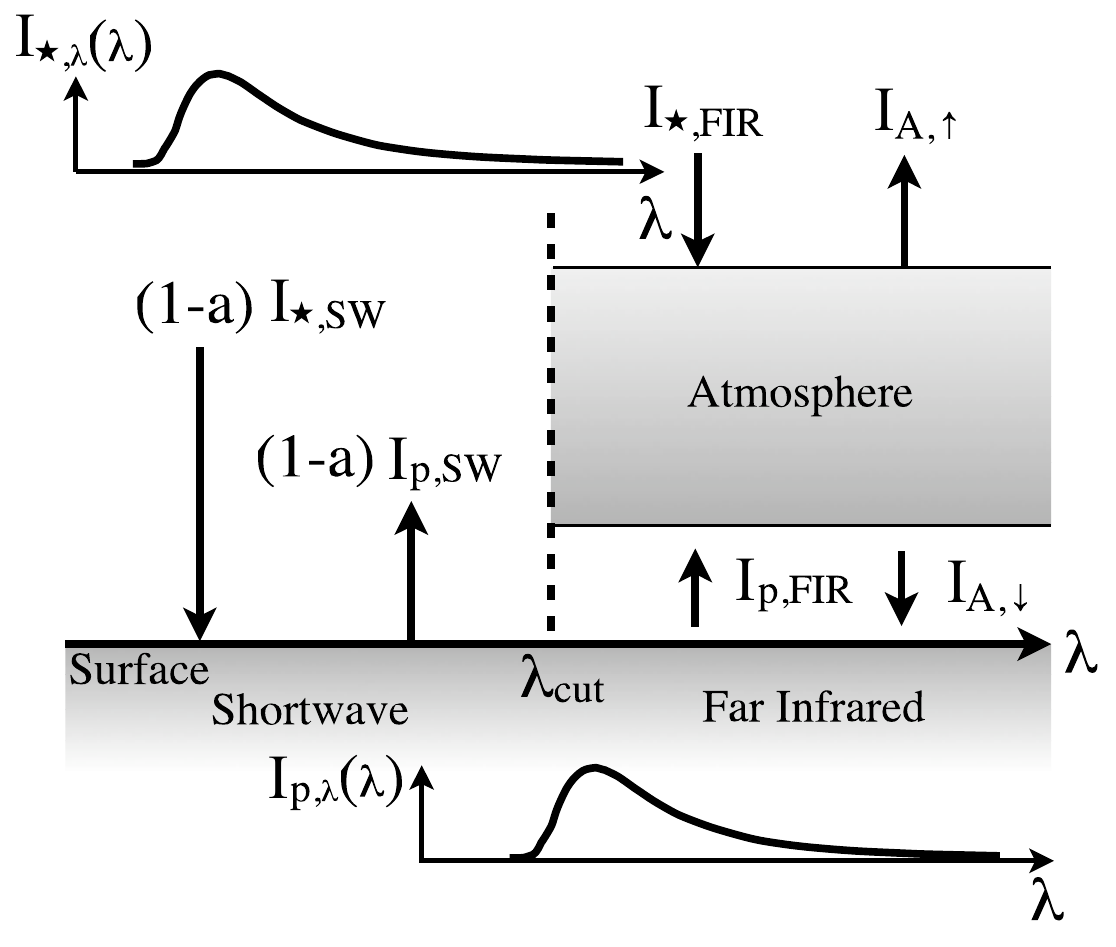,width=0.45\textwidth}} \vskip -4mm
\caption{\figstyle A heuristic description of the semi-gray model. The SW band is assumed to be optically thin. The FIR is dominated by an optically thick atmosphere. Note that we do not assume that all the solar flux is in the SW band nor that the thermal radiation of the planet is confined to the FIR.  }
\label{fig:model}
\end{figure}

To understand how saturation arises in the semi-gray model, we construct the simplest model which encapsulates the salient features. Namely, we assume an optically thin SW band and an optically thick FIR band, as described in fig.\ \ref{fig:model}.  {\oa  $\lambda_{cut}$ is the wavelength separating between the shorter and longer wavelength ranges of the absorption coefficient. In all treatments hitherto it was tacitly assumed that  $\lambda_{cut}\equiv \lambda_{rad},$ where  $\lambda_{rad}$ is the wavelength at which 
\begin{equation}
fI_{*}(T_{*},\lambda_{rad})=I_{p}(T_{p},\lambda_{rad})=g
\label{eq:rad=cut}
\end{equation}
 in the atmosphere of the planet. Here, $f$ is the attenuation factor of the stellar radiation at the planet's orbit, $T_{*}$ is the effective temperature of the star which radiates as a black-body, and $T_{p}$ is the surface temperature of the planet which also radiates like a black-body. $I_{*}$ is the stellar specific intensity at the stellar surface and $I_{p}$ is the specific intensity emitted by the surface\footnote{We denote by $I$'s either upwards or downwards flux, and by $F$ the {\em net} flux downwards. If a wavelength is given, it is per unit wavelength, otherwise, it is integrated over a whole band.}. In the case of the present Earth with $T_p=288K$ and atmospheric  composition, $g$ becomes negligible at $\lambda=\lambda_{rad}=5\times 10^4{\rm \AA},$ but this is a particular case which quickly becomes invalid as $T_p$, or $T_{*}$, or $f$, change.  The approximate equality in the case of the Earth, between the wavelength separating the two absorption domains of the microscopic physics and the wavelength separating the two domains of stellar and planetary dominant radiation is accidental, and need not be the general case. Here, we alleviate this particular tacitly assumed equality and discuss the general case in which  $\lambda_{cut}\ne \lambda_{rad}.$}{}

To solve analytically the radiative transfer through the atmosphere, we shall assume for simplicity that the atmosphere includes only scattering (i.e., ``infrared clouds"), and that it can be described with the two stream  approximation. It also implies that the temperature structure of the atmosphere is unimportant. This will allow for an analytic solution of the radiative transfer, since without thermal redistribution, the fraction going through different bands remains the same at different heights. The opposite limit, of pure absorption, would give similar results but not identical, because of the frequency redistribution when emitting at different temperatures. The full numerical solution which follows in \S\ref{sec:numerical}, solves the more complicated pure absorption problem (the algorithm of which is described in Appendix B with and without scattering).   

{\oa  Let $\tau $ be the optical depth in the FIR, measured from the top of the atmosphere downward.}{} If we write $I_+(\tau)$ as the thermal FIR flux towards larger optical depths (downwards) and $I_-(\tau)$ as the flux towards smaller optical depths, then the solutions for $I_{\pm}(\tau)$ under the above approximations are (cf. Appendix B):
\def\taut{\tau_\mathrm{tot}  }
\begin{eqnarray}
F_{\FIR}(\tau) &\equiv& I_-(\tau)-  I_+(\tau) = \mathrm{const.} \\
2J_{\FIR}(\tau) &\equiv& I_-(\tau)+  I_+(\tau) = \left[ I_-(\tau) - I_+(\tau)\right] \tau + \mathrm{const.}  \nonumber
\end{eqnarray}
or by comparing the conditions at the top ($\tau=0$) to the bottom ($\tau= \taut$), we obtain
\begin{eqnarray}
\label{eqs:two_beam}
I_-(\taut)-  I_+(\taut) &=& I_-(0)-  I_+(0) , \\
I_-(\taut)+  I_+(\taut) &=& \left[ I_-(0) -  I_+(0)\right] \taut \nonumber \\ 
& &+ \left[ I_-(0)+  I_+(0)\right] \nonumber.
\end{eqnarray}

The boundary conditions we have are:
\begin{equation}
\label{eqs:boundary} 
I_+(0) = I_{\star,\FIR}~~~{\mathrm{and}}~~~ I_-(\taut) = I_{p,\FIR} ,
\end{equation}
{\oa  where $I_{*,\FIR}$ is the insolation for $\lambda > \lambda_{cut}$ and $I_{p,\FIR} $ is the emission from the planetary surface at  $\lambda > \lambda_{cut}$. It is generally assumed that $I_{*,\FIR}=0$ and that $I_{p,\FIR}$ is the {\em total} thermal emission, an approximation which holds for the Earth at ${\rm T}=288{
\rm K}$. However, if $\lambda_{cut}$ decreases, these two assumptions are no longer justified.

The last equation is obtained from the assumption of thermal equilibrium, i.e., the total absorption equals the total emission:
\begin{equation}
(1-a)I_{\star,\SW} + I_{A,\downarrow} = (1-a) I_{p,\SW} + I_{p,\FIR},
\end{equation}
{where $I_{*,SW}$ is the insolation for $\lambda<\lambda_{cut}$, $I_{A,\downarrow}$ is the atmospheric emission  towards the surface (at $\lambda > \lambda_{cut}$), $I_{p,SW}$ the planetary emission at $\lambda < \lambda_{cut}$ and $a$ is the albedo at the short wavelengths. Our main point is that $I_{p,SW}$ must be included in the energy balance at relatively high surface temperatures. }

Using the two sets of eqs. \ref{eqs:two_beam} and  \ref{eqs:boundary}, the thermal equilibrium becomes:
\begin{equation}
\label{eq:twobeam_eq}
(1-a) \left( I_{\star,\SW} - I_{p,\SW} \right) =  { 2 (I_{p,\FIR} - I_{\star,\FIR}) \over 2 + \taut}.
\end{equation}

The behavior for $\taut \rightarrow 0$ is as expected, giving direct equilibrium between the incoming stellar flux and the outgoing thermal radiation. However, the ``Simpson limit" of large optical depths, together with $I_{\star,\FIR} = 0$ and $I_{p,\SW} =0$, gives that $(1-a) I_{\star,\SW} \approx I_{p,\FIR} / (\tau/2)$. This is an artifact of the two stream  approximation. In the full radiative transfer solution for this limit, one should obtain $(1-a) I_{\star,\SW} \approx I_{p,\FIR} / (3 \tau/4)$. Thus, we should modify eq. \ref{eq:twobeam_eq} to be  
\begin{equation}
\label{eq:full_eq}
(1-a) \left[ I_{\star,\SW} - I_{p,\SW}(T_p) \right] =  {  \left[I_{p,\FIR}(T_p) - I_{\star,\FIR}\right] \over 1 +  3\taut/4}.
\end{equation}
This equation is essentially an equation for $T_p$. 

We assume that both the stellar and planetary emissions can be adequately described as black bodies.
The stellar radiation reaching the surface of the planet in the shortwave is given by
\begin{equation}
I_\mathrm{\star,SW}={1\over 4}\left( {R_{\star}^2 \over d^2 }\right) \int_{\ncut}^\infty \pi B(T_{\star},\nu ) d\nu.
\end{equation}
Here $d$ is the distance of the planet from the star. The factor 4 arises from the ratio between the total surface area of the planet to its cross-section to the stellar radiation assuming fast rotation. For convenience we work in frequency space, and with $\ncut \equiv c/\lcut$.

The  long wavelength stellar energy flux at the top of the atmosphere is given by
\begin{equation}
I_\mathrm{\star,FIR}={1\over 4} \left( {R_{\star}^2 \over d^2 }\right) \int_0^{\ncut } \pi B(T_{\star},\nu ) d\nu .
\end{equation}

The emission of the planet at short wavelengths is given by:
\begin{equation}
I_\mathrm{p,SW}=\int_{ \ncut }^\infty \pi B(T_{p},\nu ) d \nu ,
\end{equation}
and at long wavelengths by 
\begin{equation}
I_\mathrm{p,FIR}=\int_0^{\ncut} \pi B(T_{p},\nu ) d \nu .
\end{equation}

If we define $f \equiv (R_\star/d)^2/4$ and plug the fluxes into the surface equilibrium given by eq.\ \ref{eq:full_eq}, we obtain an equation for the equilibrium temperature $T_p$ as a function of the stellar emission, the distance to it, the optical depth in the IR, and of $\ncut$:
\begin{eqnarray}
\label{eq:full_equation}
 (1-a)  \int_{\ncut}^\infty \left[ f B(T_{\star},\nu ) - B(T_{p},\nu )  \right] d\nu  \hskip 1.5cm  &&  \\ \nonumber
 = {1 \over 1 +  3\taut/4}  \int_0^{\ncut} \left[   B(T_{p},\nu ) - f B(T_{\star},\nu )  \right] d\nu.    && 
 \end{eqnarray}

\section{The ``Simpson" Solution}
\label{sec:simpson}

To understand the Simpson paradox, we begin by imposing the usual approximations in eq. \ref{eq:full_equation}. The ``Simpson solution" is obtained if the planetary SW emission is neglected on the l.h.s., while the stellar component in the FIR is neglected on the r.h.s.\ of the governing  equation \ref{eq:full_equation}. Furthermore, one also assumes that $\ncut$ is well below the incoming shortwave emission but well above the outgoing planetary radiation. Under the latter approximation, the integrals give the Stephan-Boltzmann emission, namely:  $\int_0^\infty B(T,\nu) \nu = \sigma T^4/\pi $ and eq.\ \ref{eq:full_equation} reduces to
\begin{equation}
T_p \approx  \left[ \left( 1+{3 \tau_{tot}\over 4}\right) {(1-a) \over 4} \left( R_\star \over d \right)^2 \right]^{1/4} T_\star .
\end{equation} 

This result demonstrates how the planetary temperature diverges as $\propto \tau_{tot}^{1/4}$. The main reason for the divergence is the fact that irrespective of how high the planetary temperature is, there is no route for the planetary radiation to escape except through the optically thick atmosphere.

{\oa  All other radiative transfer solutions found in the literature to overcome the greenhouse runaway involve a window in the IR due to water vapors (or any other species).
The exception is \cite{Kasting88} who suggested that leakage through the SW can in principle serve as such a route. 
 }{}
 
Besides radiative transfer, another physical mechanism that can advect the energy from the surface towards higher up in the atmosphere is that of convection. If it arises, it will necessarily reduce the equilibrium temperature, and it therefore implies that any solution we later find is necessarily an upper limit. Some authors concluded that convection may become important in a runaway planet, especially if it is dominated by water \citep[e.g.,][]{Kasting88}. Others, on the other hand, found that it is not \citep[e.g.,][]{Matsui}. In the only example we know, that of Venus, most of the atmosphere is not convective. 

In this respect, is also worth mentioning that that unlike convection, the radiative solution is insensitive to the physical length scale. This implies that a given  radiative solution may be adequate to describe one planet, but convection may appear on another with a similar atmosphere and energy budget.Ó  
 
\section{Adding a window in the IR}
\label{sec:IRwindow}

{\oa Although Simpson realized the existence of a paradox, he  argued that the water vapor window should  eliminate it \citep{Simpson2} but without  evaluating the impact of this assumption on his paradox.
 \cite{Weaver}    considered the effect of such a window in mitigating the runaway and even harnessing it.  They  applied  the semi-gray model, as assumed by Simpson, and added  a transparent window to the old Schwarzschild problem. They allowed the scale height of the trace greenhouse gas to be different from the one of the main constituents and accounted for strong non-overlapping lines within the context of semi-gray optical depth definition.  Weaver and Ramanathan assumed that the short wavelengths range is   completely transparent and solved for the IR flux only. Adopting  the two stream approximation they wrote that:

 \begin{equation} {dI_+ \over d \tau}=D(I_+ -\pi B)~~~and ~~~{dI_-\over d\tau}=-D(I_- -\pi B), \end{equation}
 where $D$ is a constant that arises from the integration over a hemisphere, and $\tau$ is the constant optical depth in the IR. Usually $D=3/2$. 
 
 For radiative equilibrium, the authors imposed the condition: $dF/d\tau=0,$ where $F=I_+-I_-=- F_0$. Here $F_0$ is 
 incoming solar radiation reaching the surface, which has to be radiated upwards in the FIR, thereby giving a negative net flux $F$. 
 For global mean condition for $F_0$ is given by 
 \begin{equation}  F_0={S\over 4}(1-a), \end{equation}
 where  $S$ is the solar insolation. The boundary conditions are:
 \begin{equation}  
 I_+(\tau=0)=0 ~~~~~~~ \mathrm{and} ~~~~~~~I_-(\tau)=\pi B_g=I_+(\tau)+F_0. 
 \end{equation}
 The solution for the temperature is:
 \begin{equation}  \sigma T^4=F_0{(1+D\tau)\over 2}~~~~~ \mathrm{and} ~~~~~\sigma T_p^4=F_0{(2+D\tau_{tot})\over 2}. 
 \label{eq:WRsol}
 \end{equation}
 Note that there is a discontinuity between the surface and the atmospheric temperature. 

 What happens when the IR domain contains a window?  Weaver and Ramanathan defined a parameter $\beta$ as follows:
 \begin{equation}
 \beta ={\int_{\lambda_1}^{\lambda_2}B_{\lambda}(T)dT \over \int_{0}^{\infty}B_{\lambda}(T)dT}
 \end{equation}
where $(\lambda_1 , \lambda_2)$ is the wavelength interval of the window. 
 
 The solution for the radiation field  is then:
 \begin{equation} \sigma T^4(\tau)=F_0{(1+D\tau )\over (2+\beta D \tau_{tot})}
 \end{equation}
The surface temperature again has a discontinuity, and it is given by:
 \begin{equation} \sigma T_p^4(\tau_{tot})=F_0{(2+D\tau_{tot} )\over (2+\beta D \tau_{tot})}.
 \end{equation}
 In the limit of $\beta\rightarrow 0$, we recover Simpson's solution with its paradox, namely, $T_p\rightarrow \infty$ with $\tau_{tot}^{*}\rightarrow \infty$. However, for a finite $\beta$, there is no divergence in the limit $\tau_{tot} \rightarrow \infty$. Instead, the surface temperature becomes independent of $\tau$ and tends to:
 \begin{equation}\lim_{\tau_{tot}\rightarrow \infty} \sigma T_p^4 = {F_0 \over \beta}.
 \end{equation}
 
  According to the authors,  $\beta$ is independent of temperature, and the result should be compared with eq.\ \ref{eq:WRsol}.
  However, this is not the case.
  
  Consider three typical windows in the IR. The first window is at wavelength $(2\times 10^4{\rm \AA},3\times 10^4 {\rm \AA} )$, the second at $(5\times 10^4 {\rm \AA},8\times 10^4 {\rm \AA} )$ and the third at $(10^5{\rm \AA},1.5\times 10^5{\rm \AA} )$. Their corresponding $\beta$ is depicted in fig.\ \ref{fig:window}.

\begin{figure}
\begin{center}
\epsfig{file=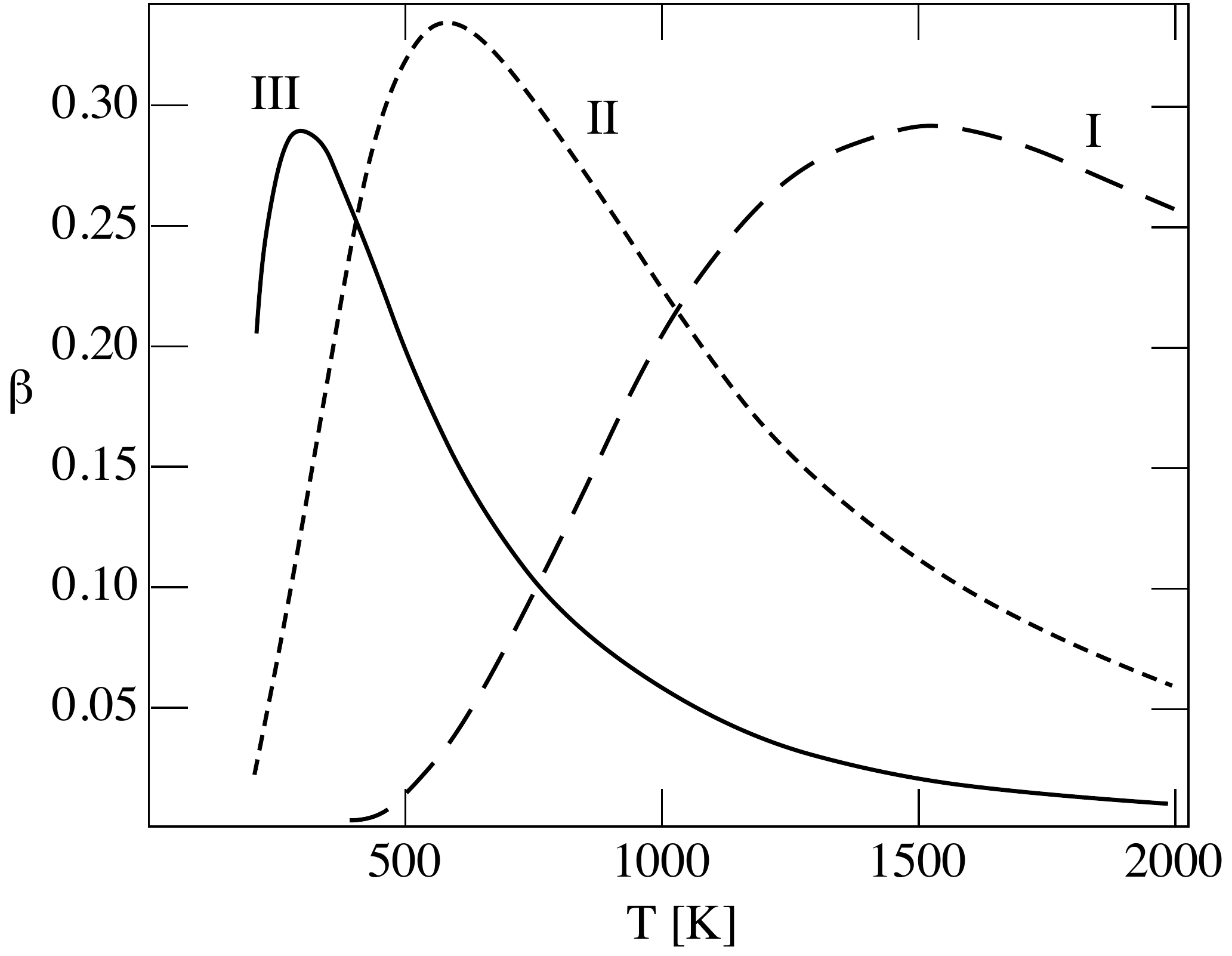, width=0.45\textwidth} \vskip -6mm
\caption{  {\figstyle  The behavior of $\beta (T)$ as a function of temperature, for three typical windows. The first window is $(2\times 10^4{\rm \AA},3\times 10^4{\rm \AA})$, the second is at $(5\times 10^4 {\rm \AA},8\times 10^4{\rm \AA})$ and the third is at $(1\times 10^5{\rm \AA},
1.5\times 10^5{\rm \AA})$}.}    \vskip -4mm
\label{fig:window}
\end{center}
\end{figure}

We find that $\beta$ is not  temperature independent. Instead, $\beta (T) $ first increases with temperature (and lowers the greenhouse temperature) and then decreases towards zero (and raises the greenhouse temperature). Hence, we conclude that a window cannot resolve the Simpson paradox  by preventing the temperature divergence.

  \cite{Pujol03} introduced a semi-gray model into  radiative-convective 1D model with a stratosphere in radiative equilibrium and a troposphere fully saturated with water vapors.    They applied the formalism of Weaver and Ramanathan and defined a similar $\beta$. Their eq. (12) is essentially the result of Weaver and Ramanathan including the particular assumption on the absorption. The authors argued correctly that $\beta$  depends on temperature.

    The first conclusion reached by these authors   was that: ``The single (absolute) SKI limit found in gray atmospheres is not obtained in non-gray atmospheres with fully transparent infrared regions". In other words ``the long-wave radiation emitted by any non-gray atmosphere with fully transparent infrared regions is not bounded".  This is a consequence of not noticing that $\beta \rightarrow 0$ as $T_p$ insceases.}

\section{An overall view  for a scattering atmosphere}
We now turn to solve the semi-gray model (eq.\ \ref{eq:full_equation}), in particular, while considering that the planet's thermal radiation can escape through the SW range, and some of the stellar SW radiation is absorbed at the top of the atmosphere.   
Since there is no analytic solution to eq.\ \ref{eq:full_equation}, we solve it numerically. The results shown here are those obtained for a Main Sequence star radiating like a black body at $5800$K and a fast rotating planet (namely, one with a uniform temperature distribution) placed at the orbit of Venus, having a similar albedo. 

Fig.\ \ref{fig:TvsLambda} depicts the equilibrium {\oa surface}{} temperature as a function of $\lcut$ for different optical depths. Several points are evident. First, for each optical depth, there is an ``optimal" $\lcut$ for which the greenhouse effect is the largest. Second, the equilibrium temperature is a monotonically increasing function of the optical depth. Third, for a given $\lcut$, there is saturation of the greenhouse effect. Fourth, a larger $\lambda_{cut}>10^4{\rm \AA}$ implies a smaller $\tau_{tot}$ for which saturation is reached. Finally, for $\lambda \sim 10^4{\rm \AA}$, the largest $\tau_{tot}$ is needed to reach saturation.

\begin{figure} 
\center{\epsfig{file=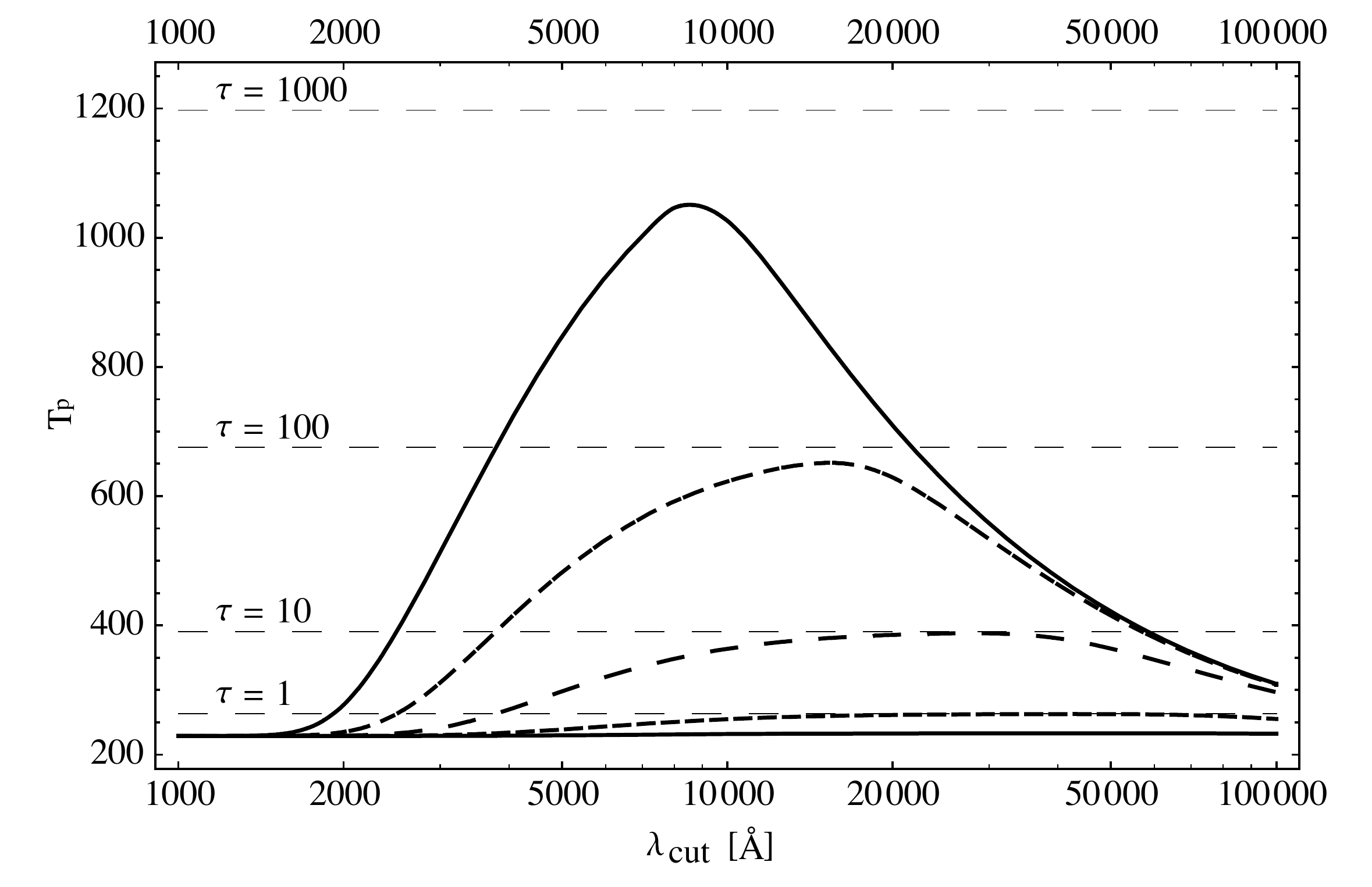,width=0.5\textwidth  }} \vskip -4mm
\caption{\figstyle The equilibrium surface temperature (eq.\ \ref{eq:full_equation}), as a function of $\lcut$ for several optical depths, $\tau_{\FIR} = 0.1,10,100,10^3$ from bottom to top. The conditions are those for a planet which has an albedo of 0.75, located  $1.08 \times 10^{13}$cm from a sun like star. The straight dashed lines denote the values expected from the Simpson solution for each optical depth. The latter solution breaks down for large optical depths.   
 }  \vskip -4mm
\label{fig:TvsLambda}
\end{figure}

Fig.\ \ref{fig:2DNoWindow} presents the equilibrium temperature in the $\lambda_{cut} - \log(\tau_{\FIR})$ plane, in which three regions can be identified.  For low $\tau_{\FIR}$ and intermediate values of $\lcut$ , the Simpson's solution is adequate. In this region, most of the flux leaks out of the planet through the moderately thick FIR. Because of this, the actual value of $\lambda_{cut}$ is irrelevant.

\begin{figure} 
\center{\vskip -15mm \epsfig{file=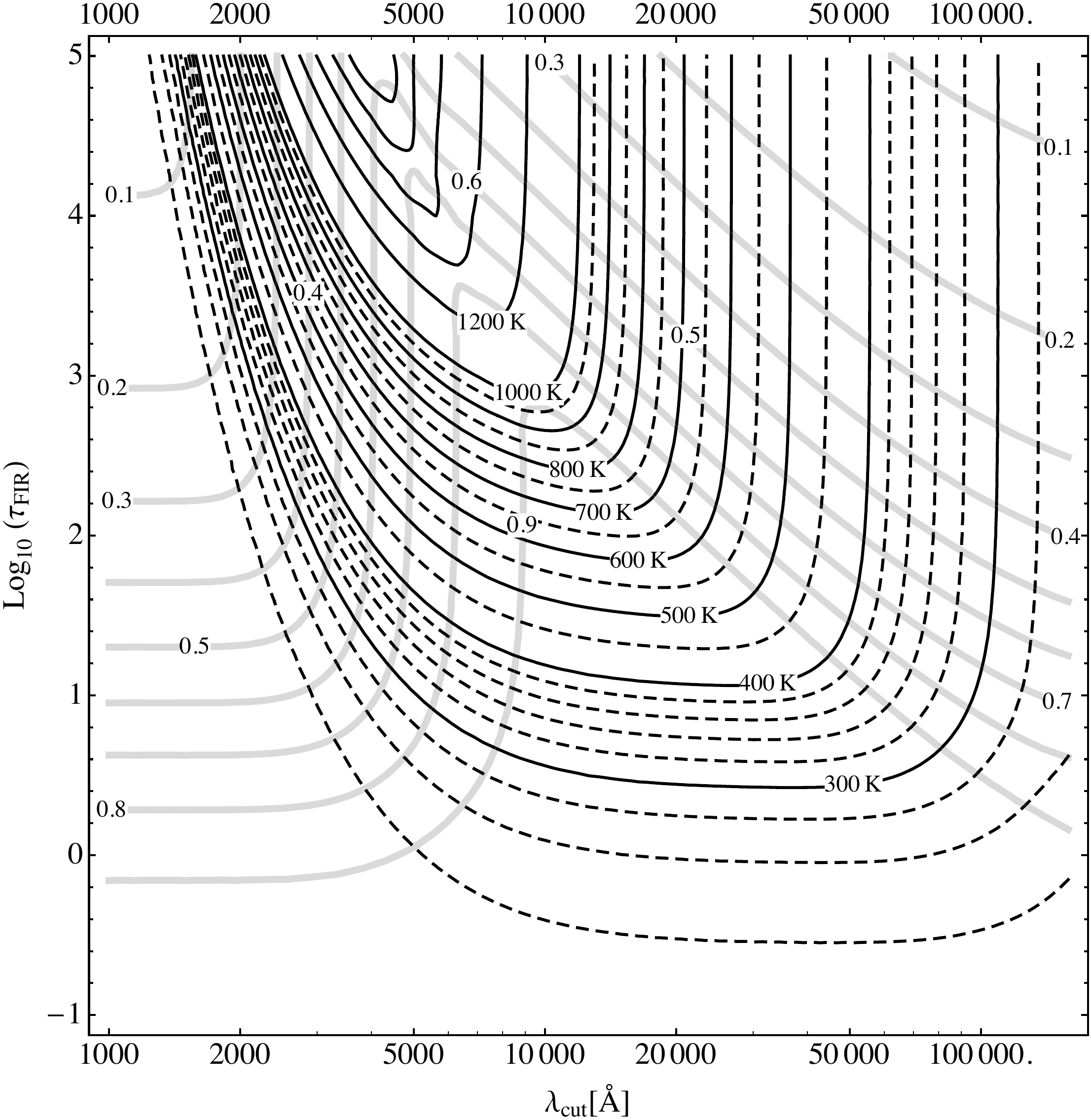,width=0.45\textwidth  }} \vskip -4mm
\caption{\figstyle The equilibrium temperature (eq.\ \ref{eq:full_equation}) as a function of $\lcut$ and $\tau_{\FIR}$, for Venus like conditions (given in fig.\ \ref{fig:TvsLambda}). Black lines denote the planetary temperature while the gray lines, the ratio between the solution temperature and the Simpson temperature. The latter is an adequate solution only for moderate optical depths and moderate cutoff frequencies. If the optical depth is too large (upper right regions), the Simpson solution breaks down because radiation escapes through the optical. If the optical is blocked (left region), less stellar radiation can reach the surface, and the Simpson solution breaks down as well.   
 }
\label{fig:2DNoWindow}
\end{figure}

If the optical depth becomes {\oa sufficiently}
  large, the temperature increases to such high values that a significant part of the  flux can leak through the SW range, below $\lcut$. For larger $\lcut$'s, in cases that the stellar IR radiation dominates that of the planet at even longer wavelengths, it is easier for the radiation to leak through the SW (which now contains longer wavelengths), and therefore saturation takes place at lower optical depths. In the $\lcut - \tau_\FIR$ plane, this takes place in the upper-right region. Once leakage through the SW becomes important, the value of $\tau_\FIR$ is unimportant (provided that it is large enough) since the small amount of radiation passing through the optically thick FIR is irrelevant

The temperature decrease on the l.h.s.\ of the plane with the decrease of $\lcut$ is a consequence of another effect. If $\lcut$ is {\oa sufficiently}   small, the atmosphere blocks some of the incoming SW radiation from reaching the surface. Instead, some of the stellar radiation will be scattered in the atmosphere and radiated back to space (through the $I_-$ component in the two stream  approximation, or more generally through $I_{A\uparrow}$, as in fig.\ \ref{fig:model}). 

Formally, for very small $\lcut$, the saturation temperature can be  quite high, though never more than $T_{*}$. However, it requires extremely large optical depths and it also assumes that there are no windows in the FIR. More realistically, at very large wavelengths beyond the rotational molecular lines, that is, beyond $\lambda \gtrsim 10^6{\rm \AA}$, the atmosphere should be optically thin. To see this effect, we plot in fig.\ \ref{fig:2DwithWindow} the equilibrium temperature obtained once we allow the thermal radiation to escape beyond $10^6$\AA. Two regions remain the same. The equilibrium temperature in the region where most of the radiation escapes through the optically thick FIR  is determined following the Simpson's solution. It is denoted here as region A. In the region denoted here as ``region B", saturation is achieved through leakage in the SW. In the third region, ``region C", saturation is achieved by allowing radiation to escape through very long wavelengths. This region is absent in fig.\ \ref{fig:2DNoWindow} where there is no long wavelength window.

One interesting aspect about the latter region is the fact that SW radiation is blocked, but radiation can escape through very long wavelengths. This gives rise to an anti-greenhouse effect (e.g., as is probably present on Titan, \citealt{McKay}). 


In fig. \ref{fig:Tmax} we see  the saturation surface temperature for a planet having a vanishing albedo  and located at a distance of $1.5\times 10^{13}$cm from a main sequence star, having a surface temperature of  $T_*=5800$K. As  $\lambda_{cut}\rightarrow 0$, the surface temperature tends to the black body temperature of the star. It does not diverge irrespective of the dilution factor, as is required by thermodynamics.

Although eq.\ \ref{eq:full_equation} does not have an analytic solution, it can be solved under the approximation that $\lambda_{cut}$ resides in the Rayleigh-Jeans long wavelenth tail of the stellar radiation, and in the short wavelength cutoff of the thermal emission. This is carried out in Appendix A, where we also show that it typically under-predicts the accurate solution by only 20K to 40K.

\section{Numerical results for an absorbing atmosphere}
\label{sec:numerical}

In addition to the analytic description of the semi-gray problem, we can also solve the full problem of radiative transfer, while alleviating the assumption of a purely scattering atmosphere. 

The details of the numerical solution for the radiation field are given in appendix B.  While the optical depth previously referred to scattering, from now on it refers to absorption. Also, unless otherwise mentioned, we will consider here a Venus like planet, i.e., a planet revolving a solar like star at a distance of $1.08 \times 10^{13}$cm, but with an $a=0$ albedo.

{\oa  We begin by discussing the $\tau_{\rm FIR} \rightarrow \infty $ limit, showing again that there is a maximal possible temperature for a given $\lambda_{\rm cut}$. We then continue to the general case and study the different limits where different physical processes govern the planetary temperature. Afterwards, we explore more specific points, the effects of a FIR window and the anti-greenhouse effect. }

{\oa 
The result of the numerical solution is shown in fig.\  \ref{fig:saturation}, where the surface temperature is plotted as a function of the total optical depth in the FIR range, for several total optical depths in the visible range. The appearance of saturation for high optical depths is clear. As $\tau_{\SW}$ increases, the saturation temperature decreases for large $\tau_{\FIR}$. Moreover, the surface temperatures for small $\tau_{\FIR}$'s decrease to below the equilibrium temperature leading to an anti-greenhouse effect.

\begin{figure} 
\center{\vskip -14mm \epsfig{file=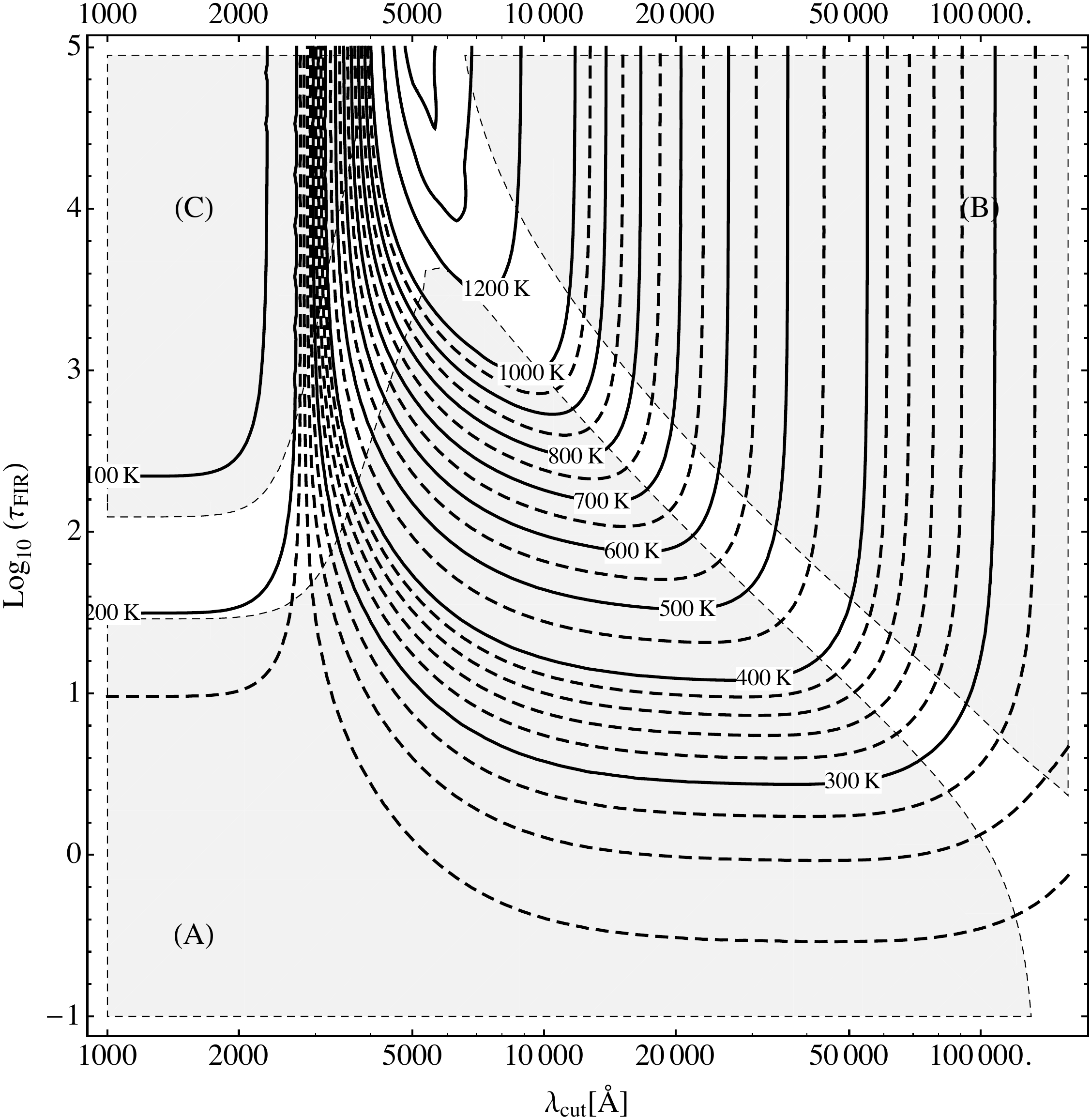,width=0.45\textwidth }} \vskip -4mm
\caption{\figstyle Similar to fig.\ \ref{fig:2DNoWindow} with the exception that the atmosphere is assumed to be optically thin for $\lambda > 10^6$\AA. The gray regions here define the characteristics of the solution. In region (A), more than half of the radiation escapes through the optically thick FIR. In region (B), more than half of the radiation escapes below $\lcut$, i.e., in the optical and NIR. In region (C), more than half of the radiation escapes through the optically thin window beyond 10$^6$\AA. Note that in this region, the system has an anti-greenhouse behavior. This is because SW radiation cannot easily penetrate the system, but sufficient radiation can escape in the  FIR ``window".  
 }
\label{fig:2DwithWindow}
\end{figure}

\begin{figure}[t]
\begin{center}
{ \vskip -4mm \epsfig{file=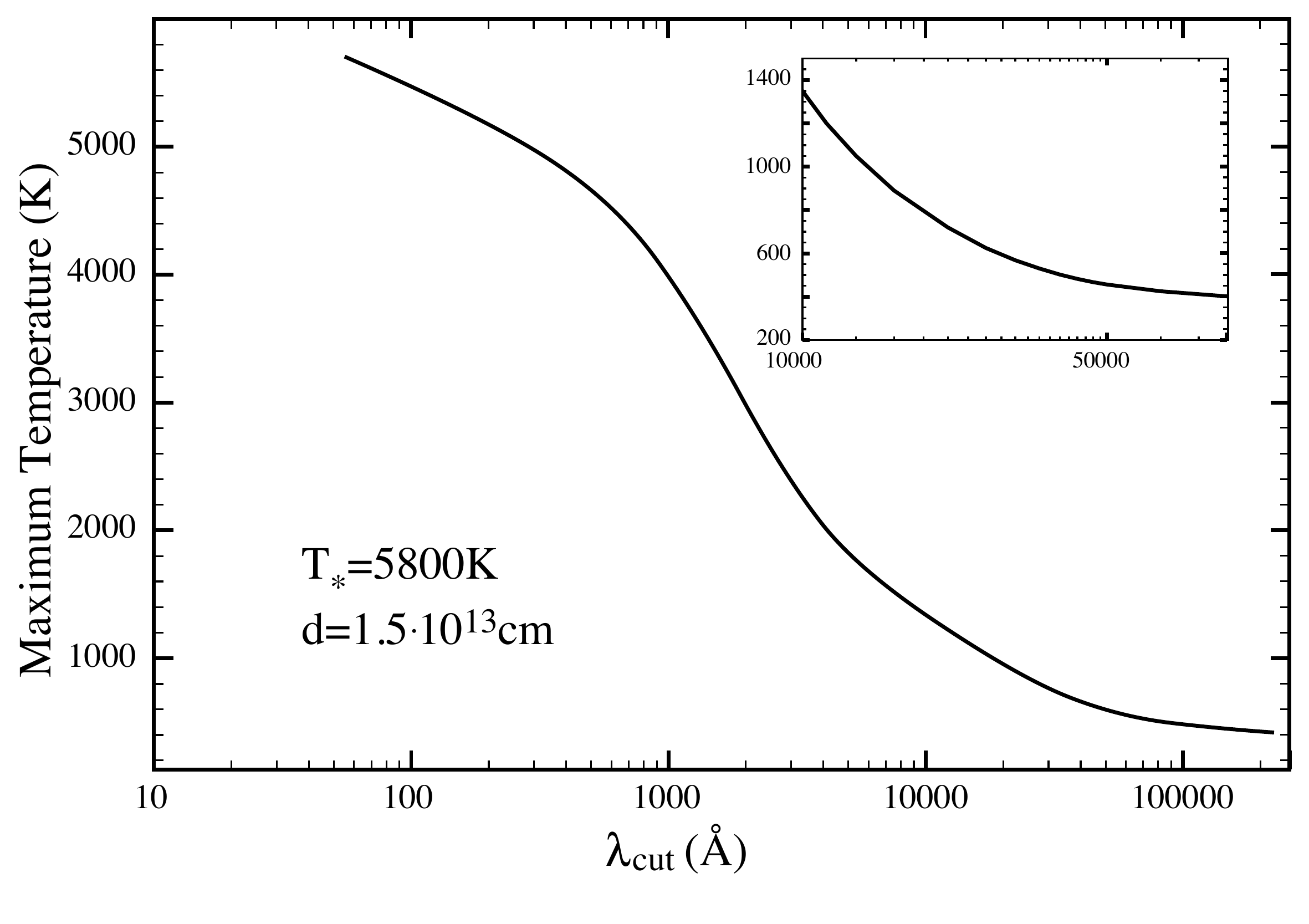, width=0.45\textwidth}} \vskip -4mm
\caption{  {\figstyle  The maximum possible greenhouse temperature (eq.\ \ref{eq:full_equation}), with $\tau_{\FIR}\rightarrow \infty$ as a function of $\lambda_{\rm cut}$, for an $a=0$ planet at a distance of $1.5\times 10^{13}$cm from a main sequence star with a surface temperature of 5800K radiating like a black-body. In principle, the planetary temperature can asymptotically reach the stellar temperature, but this requires $\lambda_{\rm cut}$ to be unrealistically small, and $\tau_{\rm FIR}$ to be unrealistically large.}}  
\label{fig:Tmax}
\end{center}
\end{figure}

\begin{figure} 
\center{ \vskip -4mm \epsfig{file=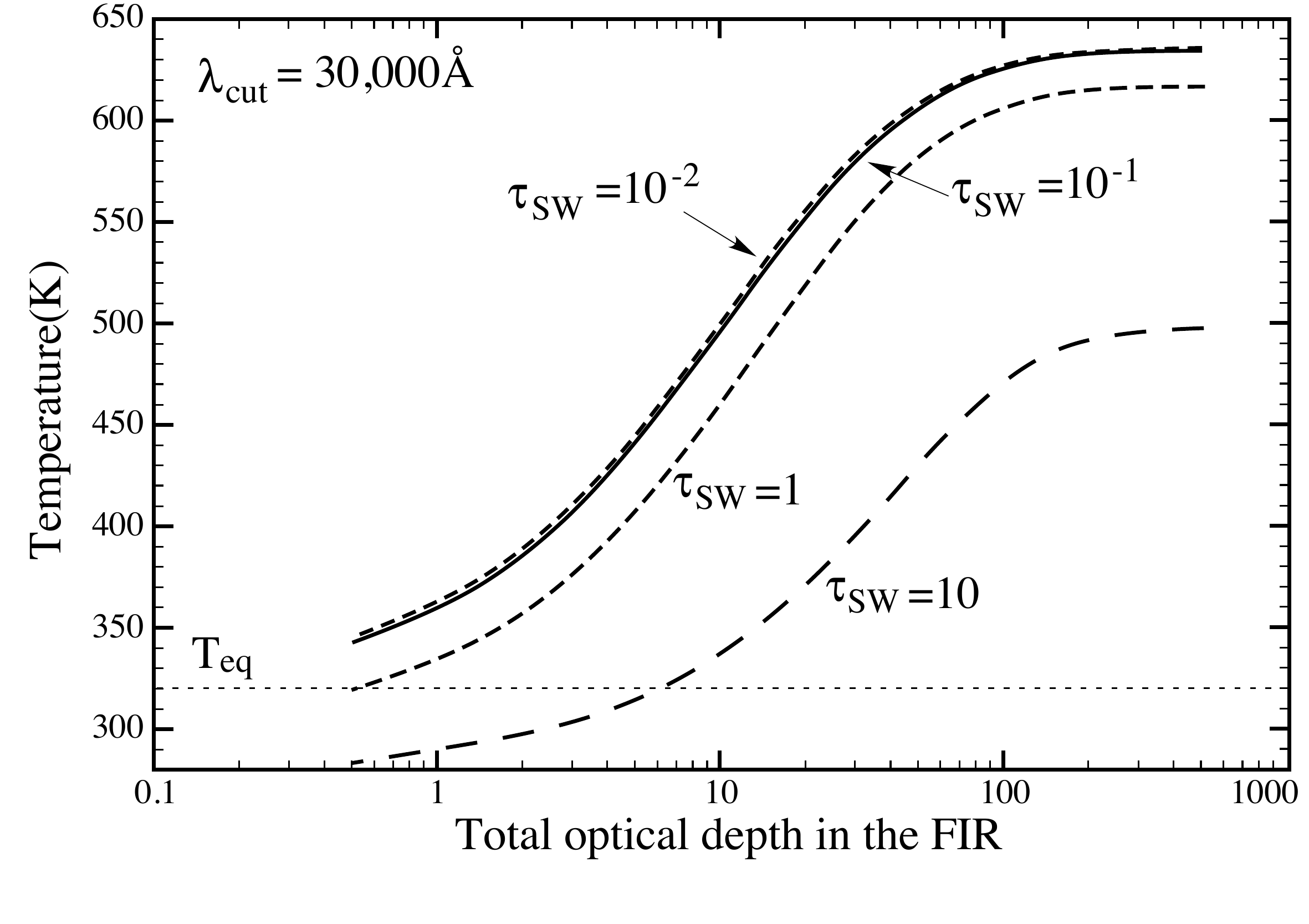,width=2.7in,width=0.45\textwidth }} \vskip -4mm
\caption{\figstyle The surface temperature as a function of the total optical depth in the FIR, calculated for several total optical depths in the visible using the full numerical calculation. The model planet is one with $a=0$, at the distance of Venus from a sun like star, with $\lambda_{\rm cut} = 30,000{\rm \AA}$. An anti-greenhouse effect is obtained for $\tau_{\rm SW}=10$ and $\tau_{\rm FIR}<6$.  }
\label{fig:saturation}
\end{figure}

The effect of $\lambda_{cut}$ is shown in fig.\ \ref{fig:Tsut(lcut)}. As $\lambda_{cut}$ decreases, the saturation temperature increases and reaches a maximum. This increase takes place because more of the thermal emission is blocked by a high FIR optical depth atmosphere. The subsequent decrease for smaller $\lambda_{cut}$ takes place because more of the stellar insolation is in the optically thick wavelength band. This radiation is absorbed at a high altitude and re-emitted to space without reaching the surface. Thus, less radiation in the visible reaches the surface.


\begin{figure}
\center{\vskip -4mm \epsfig{file=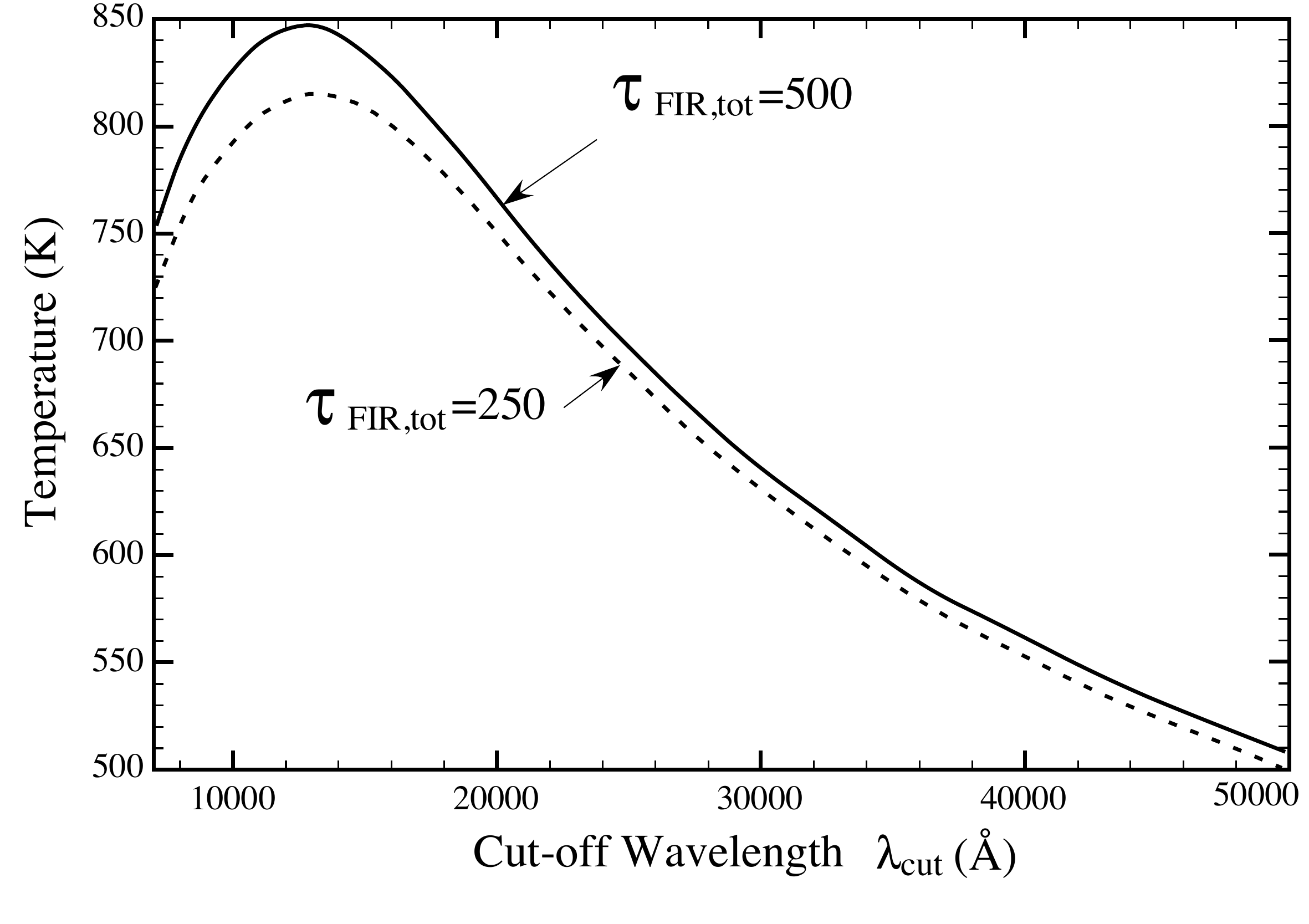,width=2.7in,width=0.45\textwidth  }} \vskip -4mm
\caption{\figstyle The variation of the saturation temperature with $\lambda_{\rm cut}$, as calculated with the full numerical calculation for the case described in fig.\ \ref{fig:saturation}, with $\tau_\SW=0.1$. The difference between the $\tau=500$ and $\tau=1000$ cases is too small to be discerned, thus demonstrating the saturation.}
\label{fig:Tsut(lcut)}
\end{figure}

}{}

{\oa 
\subsection{Saturation of the FIR emission}
Let us return to eq. \ref{eq:full_eq} and consider the limit $\tau_{tot}\rightarrow \infty$. In this limit $I_{*,SW}=I_{p,SW}(T_p)$  irrespective of the albedo. 

Furthermore, by considering that $I_{*,SW}=I_{\star}-I_{*,\FIR}$ (where $I_{\star}$ is the bolometric stellar flux) in the energy balance equation for the top of the atmosphere, one finds that
$ 
I_{*,SW}+I_{*,FIR}=I_{p,SW}+I_{A,\uparrow},
$ 
where $I_{A,\uparrow}$ is the upwards emission by the top of the atmosphere, in the FIR range. 

One therefore has in the $\tau_{tot}\rightarrow \infty$ limit that
\begin{equation}
 I_{A,\uparrow}=I_{\star}-I_{p,SW}=I_{*,FIR}.
 \end{equation}
This implies that the emission in the FIR from a runaway greenhouse planet is simply the ``reflected" stellar component in the FIR, which is less than the value expected for a small $\tau_\FIR$. In the latter case, the FIR includes also the ``reprocessed"  stellar SW radiation. This may be somewhat counter intuitive, that a planet which has a strong greenhouse actually has a smaller emission in the FIR. 

Let us now study numerically the FIR behavior of a planet with a strong greenhouse effect.  Fig.\ \ref{fig:FIRsaturation15} depicts  the fraction of the radiation  emitted to space from the top of the atmosphere in the FIR, as a function of the total optical depth in the FIR range, for several cases of $\tauSW$. We observe that as $\tau_{\FIR}$ increases, which also increases the surface temperature, the fraction of the radiation emitted above the cut-off decreases and reaches an asymptotic value. This is consistent with the above argumentation. 


\begin{figure} 
\center{\epsfig{file=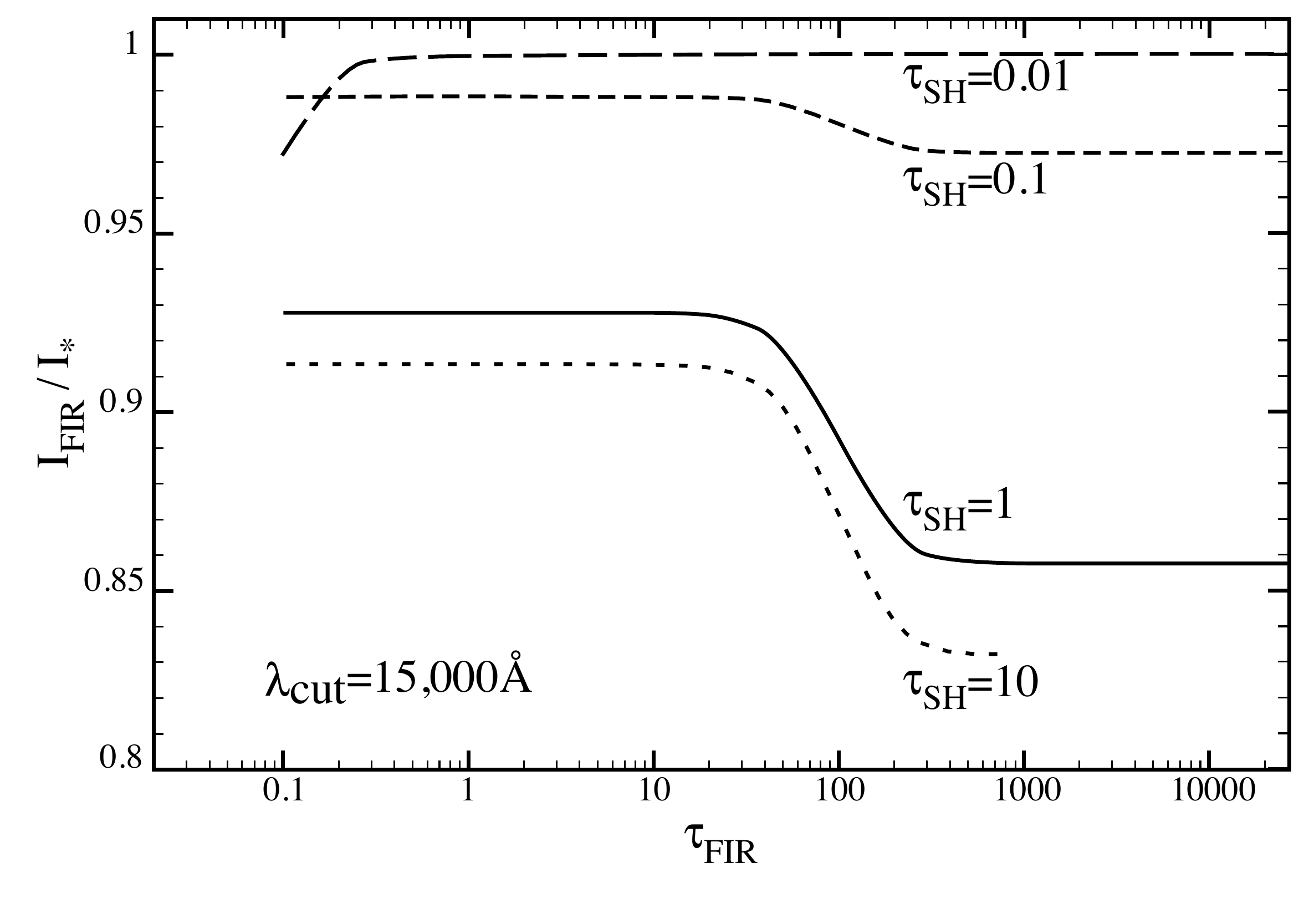,width=0.45\textwidth  }} \vskip -4mm
\caption{\figstyle The fraction of the flux coming out in the FIR as a function of the total optical depth in the  short wavelength range, $\lambda<\lambda_{cut}$, as obtained with the full numerical calculation. 
 }
\label{fig:FIRsaturation15}
\end{figure}

Next, let us consider the dependence of $I_{\FIR}/I_\star$ as a function of $\eta\equiv\kappa_{\FIR}/\kappa_{\SW}$, as shown in fig.\ \ref{fig:emitted-flux}. We observe that for   $\eta \gg 1$ the emitted radiation obtains an asymptotic value which depends on $\tau_\SW$. Moreover, as $\eta$ decreases, $I_{\FIR}/I_\star$ reaches a maximum as $\eta$ decreases, for $\eta \sim 1$.

\begin{figure}[t] 
\center{\epsfig{file=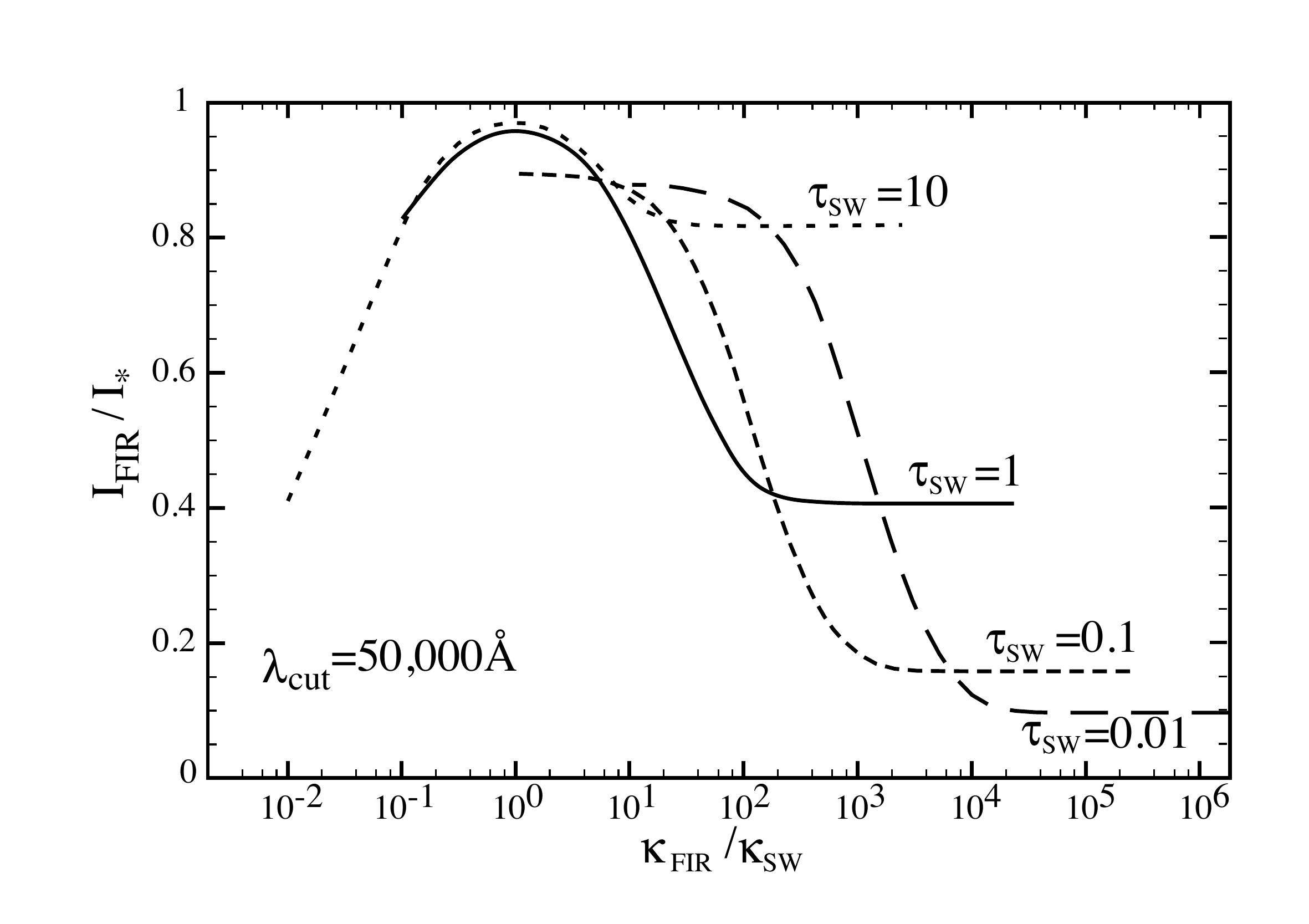,width=0.45\textwidth  }} \vskip -4mm
\caption{\figstyle The same as figure \ref{fig:FIRsaturation15} but for $\lambda_{cut}=5\times 10^4{\rm \AA}$ and as a function of 
$\eta=\kappa_{\FIR}/\kappa_{\SW}$.  $I_{\FIR}=I(\lambda>\lambda_\mathrm{cut})$ is the FIR radiation emitted from the star. }
\label{fig:emitted-flux}
\end{figure}

We show in fig.\ \ref{fig:IFIR(Tp)}  how the FIR emission varies with the surface temperature. The observed behavior is far from trivial, even the sign of the dependence of $I_\FIR$ with respect to the surface temperature changes!

%
}{}

\begin{figure} 
\center{\epsfig{file=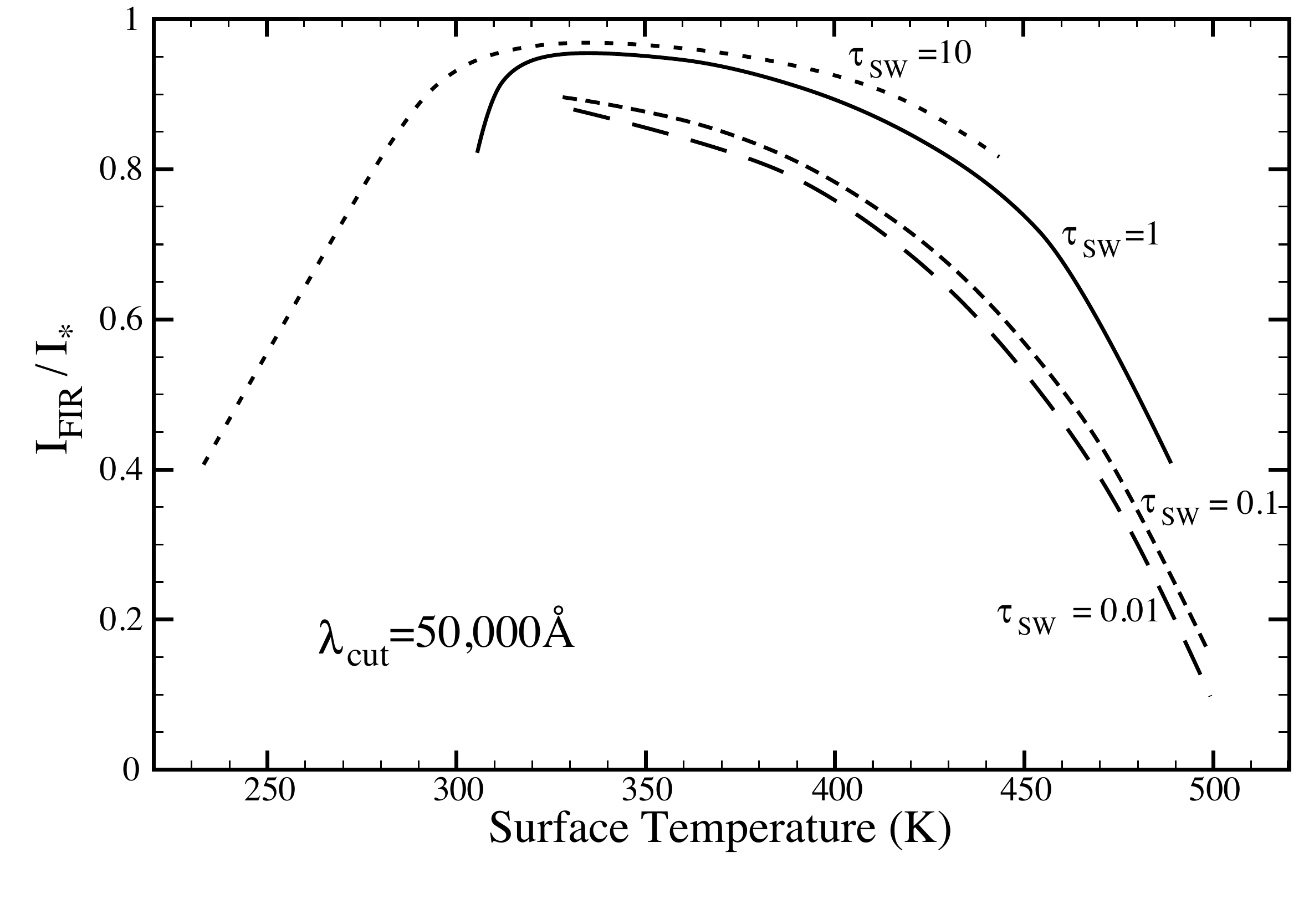,width=2.7in,width=0.45\textwidth }}
\caption{\figstyle The same as figure \ref{fig:FIRsaturation15}, but as a function of the surface temperature.  }
\label{fig:IFIR(Tp)}
\end{figure}

\subsection{The anti-greenhouse effect}
\label{sec:anti-green}
{\oa  

Let us return to fig.\ \ref{fig:IFIR(Tp)}.  As $\tauSW$ increases to about unity and higher, the surface temperature decreases to below $T_\mathrm{eq}$ (obtained for vanishing $\tau_{SW}$ and $\tau_\FIR$, that is, assuming no atmosphere, which for the depicted case is about 325K). The reason is as follows. When $\tauSW$ increases to about unity, most of the incident radiation is absorbed at low optical depths (high altitudes) and as a consequence the temperature saturates, or even decreases inwards. As $\tauSW$ increases, less radiation reaches the surface and the surface is therefore required to emit in equilibrium, less radiation.  However, if $\tau_\mathrm{FIR}$ is smaller than $\tau_\mathrm{SW}$, the smaller required emission from the surface translates into a reduced temperature. 

 In fact, an optical depth of $\tauSW \approx 10^{-2}$ is sufficient to heat the top of the atmosphere so as to invert the temperature gradient. This is the anti-greenhouse effect and it appears for $\tauSW \ge 10^{-2}$ and small $\tauFIR$'s. But as $\tauFIR$ increases to about $\tauSW$ and more, the classical greenhouse effect (hot surface and negative temperature gradient with height) re-emerges.  A similar solution was found by \cite{McKay}, who used a radiation diffusion model.
 
 It is obvious that there exist for every $\lcut$ and $\tauSW$ a value of $\tau_{\FIR}$ for which the insolation results in an almost isothermal temperature gradient in the atmosphere despite the apparent heating.  We find that without exception,  
\begin{equation}
\left({\partial \Tsur \over \partial \tauSW}\right)_{\tauFIR=const} <0
\end{equation}
namely, an increase of the absorption in the short range decreases the surface temperature simply because stellar energy is absorbed higher in the atmosphere and can more easily escape from the planet.  The magnitude of the derivative depends on $\tauFIR$ and increases (in absolute value) with $\tau_{\FIR}$. 
}
\section{Application to Venus}  
\label{sec:discussion}

In relating the results of the idealized model to a real system,  we have to be aware that the relevant choice of $\lcut$ and $\tau_{\FIR}$ depends on the actual  constituents  of the planetary atmosphere. Moreover, we also have to relate the real line opacity to the effective wide-band absorption assumed here, for which  an appropriate averaging is required. 

From eq.\ \ref{eq:full_eq}, we can write an expression for the average net FIR flux (per unit frequency) over a finite band $\Delta \nu$, and define an effective opacity through the following:
\begin{eqnarray}
\overline{\Delta F}_{\FIR}  &=& {1 \over \Delta \nu}  \int_{\nu_1}^{\nu_2} 
{ { \left[F_{p,\FIR}(T_p) - F_{\star,\FIR}\right] \over 1 +  3\tau(\nu) /4} } d \nu \\ &\approx &   
{{ \left[\overline{F}_{p,\FIR}(T_p) - \overline{F}_{\star,\FIR}\right] \over \Delta \nu} 
 \int_{\nu_1}^{\nu_2} 
{ d\nu \over 1 +  3\tau(\nu) /4} } \nonumber  \\
&\equiv & { \left[\overline{F}_{p,\FIR}(T_p) - \overline{F}_{\star,\FIR}\right] \over \Delta \nu} { 1 \over 1 +  3\tau_\mathrm{eff} /4} , \nonumber
\end{eqnarray}
with $\Delta \nu=\nu_2-\nu_1$, that is,
\begin{equation}
\tau_\mathrm{eff} \equiv { 4 \over 3} \left[   \Delta \nu \left( \int_{\nu_1}^{\nu_2} 
{ d\nu \over 1 +  3\tau(\nu) /4} \right)^{-1} -1\right] .
\label{eq:taueff}
\end{equation}
This average is equivalent to the Rosseland mean relevant for to the radiation transfer, except that optically thin frequency ranges do not cause the flux to diverge.

Fig.\ \ref{fig:CO2spectrum} depicts a specific example, and plots the total effective optical depth for two cases with a 20 bands approximation, at the high temperature and pressure present on Venus. In the first case, only CO$_2$ lines are considered, such that a window is left at $\sim 70$,$000$\AA. However, once additional ``trace" constituents are added, the spectral window fills, leaving an optically thick spectrum from the optical to the very far IR. For this composition, temperature and pressure, we obtain $\lcut \sim 6000$\AA\ and $\tau_\FIR \sim 10^4$. From fig.\ \ref{fig:2DwithWindow} we find that the equilibrium temperature for these values should be $T_{eq} \sim 1200$K. This is much hotter than the real surface temperature on Venus. However, the ideal semi-gray model was developed under the assumption that the shortwave band is optically thin. This however is not the case on Venus because of aerosols and clouds. The finite optical depth $\tau_{SW}$ implies that the solar radiation is deposited at different heights, and only a little at the surface itself. An effect that lowers the surface temperature.

\begin{figure} 
\center{\epsfig{file=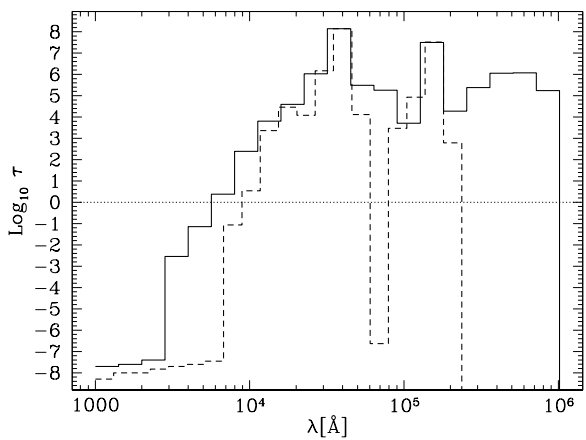,width=2.7in,angle=-90,width=0.45\textwidth  }}
\caption{\figstyle The {\em effective} band averaged optical depth (as defined in eq.\ \ref{eq:taueff}) as a function of frequency for CO$_2$ at 90 bar and 700K, and for  the same atmosphere but with added tri-atomic trace elements: 0.014\% SO$_2$ and 0.002\% H$_2$O. The line opacity is calculated using the {\sc hitran} 2004 compilation, and includes thermal and pressure broadening. Just by itself, the CO$_2$ atmosphere would have an optically thin window at $\sim 70$,$000$\AA, and beyond $\sim 250$,$000$\AA. These would allow enough radiation to escape such that  the runaway greenhouse would saturate at relatively low temperatures. With the additional tri-atom molecules present on Venus, the windows become optically thick, and $\lcut$ moves to even shorter wavelengths.  
 }
\label{fig:CO2spectrum}
\end{figure}

In fact, one can think of two limits in which the semi-gray model presents a good approximation to reality. The semi-gray model corresponds to an optically thin SW band. However, if $\tau_{SW}$   is significant, then  SW radiation cannot penetrate the atmosphere and reach the surface before being scattered. This implies that the SW energy is deposited at some finite height. Up to that height, the atmosphere should be close to isothermal while above it, one can assume that the SW band is optically thin, implying that the surface temperature should correspond to the temperature at this height, and $\tau_\FIR$ should be the optical depth in the FIR of this layer.  The case of Venus is therefore intermediate. 


\section{Summary}

Technically, Simpson's solution assumes  implicitly that the stellar radiation   penetrates the atmosphere, and consequently, the greenhouse effect is equivalent to an atmosphere illuminated in the IR but from below. The result is that  the radiation energy density at the bottom of the atmosphere grows linearly with the optical depth (in the IR), and the temperature increases like the the forth root of $\tau_{\FIR}$. We have shown that this behavior breaks down even under the semi-gray approximation. 

Using the full semi-gray model it was shown that two primary mechanisms are responsible for the saturation of the runaway greenhouse effect. The two cases depend on the value of $\lcut$. Unless $\lcut$ is small and resides in the optical region, saturation is achieved by radiating the thermal flux of the planet through the SW tail of the thermal distribution. This has an interesting observational implication: the radiation from such a planet should be skewed towards the NIR.

If $\lcut$ is very small, then the saturation of the  greenhouse could in principle be at a very high temperature, but still less than $T_{*}$. This is far from being realistic with the molecules known to exist in planets. 
The fact that the spectrum is optically thin beyond the rotational absorption lines  limits the temperature and can even give rise to an inverse-greenhouse effect.  Although windows in the FIR for smaller wavelengths could cause saturation at a lower temperature, the presence of more than one spectroscopically active specie with 3 or more atoms can  cover the spectrum efficiently.

Together, the mechanisms imply that a runaway planet like Venus cannot have a surface temperature higher than about 1200 to 1300K. This is significantly lower than some values found in the literature. If convection arises, it could reduce the maximal temperature even more.

\vskip 1cm
Regretfully,
our long time collaborator and close friend, Professor Rainer Wehrse, who actively participated in the last draft of this paper, deceased 8.12.2009.


\section*{Appendix: A simple analytic estimate of the saturation temperature}
\label{sec:approx}

The equation describing the semi-gray model, eq.\ \ref{eq:full_equation}, does not yield to an analytical solution. Nevertheless, an approximate analytical solution can be obtained when several approximations are valid. In particular, if $\lambda_{cut}$ resides in the Rayleigh-Jeans long wavelength tail of the distribution of the stellar radiation, but it also resides in the short wavelength cutoff of the thermal emission, then the Planck distribution can be approximated and an analytical solution is obtained. This can be used to demonstrate the saturation more vividly. 

Specifically, the short wavelength component of the planetary emissivity can be described by the short wavelength limit of the Planck distribution, while the long wavelength component can be described as an integral over the whole Planck distribution minus the short wave component. For the incoming stellar component, we can do the opposite. The long wavelength contribution is described with the long wavelength limit of the Planck distribution while the shortwave contribution is described by the whole Planck integral minus the long wavelength term. 
Although $\delta_1=hc/(kT_{p} \lcut )\approx 3-30$ is generally large, which allows the above approximate description of the planetary component, the appropriate dimensionless number for the shortwave stellar component, $\delta_2=hc/(kT_{\star} \lcut)\approx 0.5-2$, is not always small and the above description can fail\footnote{Note however that the peak of the Planck distribution is at $h c/\lambda \sim {\rm few}\times kT$, therefore a ratio of unity is still on the low side.}. 

In the short wavelength, we assume that 
\begin{equation}
B_\mathrm{\lambda,SW}(T)\simeq {2hc \over \lambda^3}\exp\left(-{hc \over kT\lambda}\right).
\end{equation}
In the long wavelength, we can write 
 \begin{equation}
B_\mathrm{\lambda,FIR}(T) \simeq {2 \over \lambda^2}kT.
\end{equation}
The required  integrals are therefore given by:
\begin{eqnarray}
\int_{0}^{\lcut}B_{\lambda,s}(T_{p})d\lambda &  \simeq & 2\exp\left(-{hc \over kT_{p}\lcut}\right) \left({kT_{p} \lcut \over hc}\right)   \\ \nonumber&& \times
\left({hc^2 \over \lcut^4} \right) \left( 1+ 3{kT_{p} \lcut\over hc} + \dots \right), 
\end{eqnarray}
and
\begin{equation}
\int_{\lcut}^{\infty}B_{\lambda,l}(T_{\star})d\lambda \simeq  {2kcT_{\star} \over 3 \lcut^3}.
\end{equation}
Substituting the approximations of the Planck function for the two extreme limits into eq.\ \ref{eq:full_equation} yields, after some algebra and replacing the planet's temperature $T_{p}$ by $T_{sat}$, that
\begin{eqnarray}
\left[{\pi^4 \over 15}\left(kT_\mathrm{eq} \lcut\over hc  \right)^4 -{1\over 3}\left({T_\mathrm{eq}\over T_{\star}} \right)^3\left({kT_\mathrm{eq}\lcut\over hc}  \right)\right] &=& \nonumber  \\{1 \over \tauFIR} \left( {  \pi^4 \over 15}\right)\left( { kT_{sat}  \lcut \over hc  }  \right)^4  & &  \nonumber \\
  +   \exp\left(-{hc \over kT_\mathrm{sat}\lcut}\right)\left( {kT_\mathrm{sat}\lcut\over hc}\right)\left[1+3 {kT_\mathrm{sat}\lcut \over hc}\right] & & ~~~
\end{eqnarray}
where $T_\mathrm{eq}$ is the equilibrium surface  temperature assuming no atmosphere.
Furthermore, we denote
\begin{equation}
\alpha^{-1}=\left[   { \pi^4 \over 15}   \left(  { kT_\mathrm{eq} \lcut \over hc } \right)^4  -
 {1\over 3}  \left(  { T_\mathrm{eq}\over T_{\star} }   \right)^3\left( { kT_\mathrm{eq} \lcut \over  hc } \right)   \right],
 \end{equation}
and
 \begin{equation}
x={kT_{sat}\lcut \over hc};~~~ \beta= {1 \over \tau_\mathrm{FIR,tot } } \left( { \pi^4 \over 15} \right).
\end{equation}
With these definitions, the equation becomes
\begin{equation}
{1 \over \alpha}   -  \beta x^4=\exp\left( {1 \over x}\right)x(1+3x).
\end{equation}
We expand the equation around  $1/\alpha = \exp(1/x_0)$, namely, we write $x=x_0+\delta x$, and obtain for the saturation temperature that:
\begin{eqnarray}
T_\mathrm{sat} &\approx & \left( { h c \over k \lcut }\right) \times \\  & & \left[ {1 \over \ln \alpha }-{1 \over \ln^2 \alpha }   \left(\ln \left( {1 \over \ln \alpha }\right)   +   {3 \over \ln \alpha }+{\alpha \beta \over \ln^4 \alpha }   \right) \right]. \nonumber 
\end{eqnarray}
The assumption of a small $\delta x$ implies that the first term in the square brackets is much larger than the other term (and it can be used as a ``zeroth order" approximation) . 

A comparison between the numerical and the approximate analytic model for the saturation temperature is shown in fig.\ \ref{fig:com-tsat}. The figure reveals that the analytic approximation (calculated for the case $a=0$) under-predicts the correct $T_\mathrm{sat}$ by only 20K to 40K. Most of the discrepancy can be explained through the fact that the analytical approximation neglected the stellar contribution to the FIR, which although small, does contribute some warming.

\begin{figure} 
\center{\epsfig{file=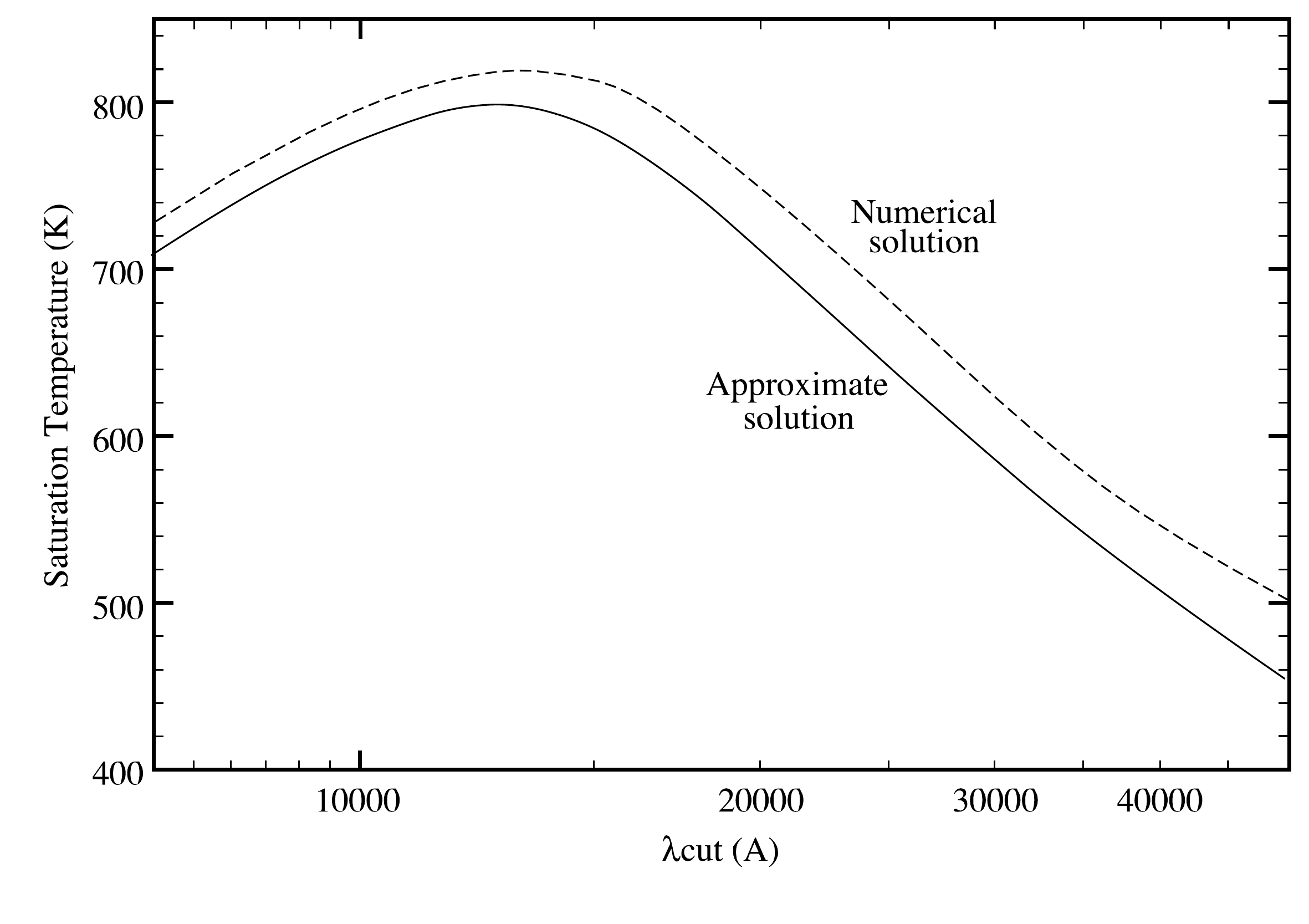,width=0.45\textwidth }}
\caption{\figstyle  A comparison between the accurate numerical solution of the radiative transfer equation (eq.\ \ref{eq:full_equation}) and the simple analytical approximation. The parameters are those of Venus but with an albedo of $a=0$. In addition $\tau_\FIR=250$.}
\label{fig:com-tsat}
\end{figure}

\section*{Appendix B: The full numerical solution}

The main results of the paper are borne from the approximate semi-gray treatment, described by eq.\ \ref{eq:full_equation}. However, since it is only an adequate description when the SW is optically thin, and it ignores thermal redistribution in the atmosphere, it is worthwhile to compare the results to the exact solution of the semi-gray problem obtained from a full numerical analysis. Here we describe the algorithm. 

\subsection*{Basic equations}

\label{set:basiceq}
In the two stream approximation, the transfer equation for the intensities of rays in the upward direction $I_+(z,\lambda )$ is
\begin{equation}
\label{transequp}
\frac{d I_+}{dz}=-\chi(z,\lambda )I_+(z,\lambda )+ \chi(z,\lambda )S(z,\lambda ),
\end{equation}
while the equation for the downward direction  $I_-(z,\lambda )$ is given by
\begin{equation}
\label{transequm}
-\frac{d I_-}{dz}=-\chi(z,\lambda )I_-(z,\lambda )+ \chi(z,\lambda )S(z,\lambda).
\end{equation}
Here the extinction coefficient $\chi(z,\lambda )$ is the sum of the absorption coefficient $\kappa(z,\lambda )$ and the scattering coefficient $\sigma(z,\lambda )$  
at height $z$ and wavelength $\lambda$. $T(z)$ is the temperature at height $z$. 

Also, the source function $S$ is given by
\begin{equation}
S(z,\lambda )=(1-\epsilon(z,\lambda ))J(z,\lambda )+\epsilon(z,\lambda )B(T(z),\lambda),
\end{equation}
with $\epsilon(z,\lambda )=\kappa(z,\lambda )/(\kappa(z,\lambda )+\sigma(z,\lambda ))$ and $B(\lambda,T)$ being the Planck function. 

The mean intensity $J$ is given by
\begin{equation}
J(z,\lambda )=\left(I_+(z,\lambda )+I_-(z,\lambda )\right)/2.
\end{equation}

$I_+$ has to fulfill the bottom boundary condition $I_+(0,\lambda )=a I_-(0,\lambda )+I_{+,\mathrm{sur}}(\lambda )$
where $a$ is the albedo of the surface and $I_{+,\mathrm{sur}}(\lambda )$ is the specific intensity emerging from the surface. The boundary condition at the top of the atmosphere is 
 $I_{-}(z_\mathrm{max},\lambda )= I_\mathrm{\star}(\lambda )$.

\subsection*{The discretized equations and their solution}
\label{dequ}

We discretize the monochromatic specific intensities and introduce the
monochromatic intensity vector.
\begin{equation}
{\bf I}(\lambda ) =(I_+(z_1,\lambda ),I_-(z_2,\lambda ),I_+(z_2,\lambda ),\dots,I_+(z_N,\lambda ))^T
\end{equation}

We describe the radiative transfer by means of transmission and reflection coefficients (cf. Shaviv \& Wehrse, 1991, Peraiah, 1984, Wehrse, Baschek \& Waldenfels, 2008) 

\begin{equation}
t=\frac{4\sqrt{\epsilon} \exp(-\sqrt{\epsilon} \tau)}
{(1+\sqrt{\epsilon})^2-(1-\sqrt{\epsilon})^2\exp(-2\sqrt{\epsilon}\tau)}
\end{equation}
and
\begin{equation}
r=\frac{(1-\epsilon)(1-\exp(-2\sqrt{\epsilon}\tau))}
{(1+\sqrt{\epsilon})^2-(1-\sqrt{\epsilon})^2\exp(-2\sqrt{\epsilon}\tau)},
\end{equation}
which are functions of the (wavelength dependent) optical depth in extinction $\tau(\lambda)$ and
  $\epsilon(\lambda)=\kappa(\lambda)/(\kappa(\lambda)+\sigma(\lambda))$.

The transfer equation for the intensity vector then reads
\begin{equation}
{\bf I} ={\bf M}{\bf I} +{\bf \Phi}{\bf B} +{\bf I}_{bc}
\end{equation}
or
\begin{equation}
{\bf I} =({\bf 1}-{\bf M})^{-1}({\bf \Phi}{\bf B} +{\bf I}_{bc})
\end{equation}
with
\begin{eqnarray}
 && \hskip -16mm
{\bf M}(\lambda ) = \\ &&  \hskip -14mm \nonumber
 \left(\begin{array}
{cccccccccc}
     0&0      &t_1(\lambda )& 0    & 0    & \dots     & 0    & 0    & 0    & 0    \\
     0& 0     &r_1(\lambda )& 0    & 0    & \dots    &  0    &  0   & 0    &  0   \\
     0&r_2 (\lambda )& 0    &  0   &t_2(\lambda )& \dots    &  0   &   0   &  0   &  0   \\
     0&t_2 (\lambda )&  0   &  0   &r_2(\lambda )& \dots    &  0   &  0    &  0   &  0   \\
\multicolumn{10}{c}{\dotfill}                                                   \\
     0& 0     &  0   &  0   & 0    & \dots     & r_N(\lambda )    &  0    & 0        &t_N(\lambda )  \\
     0& 0     &  0   &  0   & 0    & \dots     &t_N (\lambda )    &   0   &  0       &r_N(\lambda ) \\
     0&  0     & 0    &  0   & 0    & \dots     & 0      & 0     &\alpha & 0
\end{array}\right)
\end{eqnarray}
and $\alpha$ being the albedo of the surface, and
\def\mv{\hskip -1mm}
\begin{eqnarray}
 && \hskip -16mm
 {\bf \Phi}(\lambda ) = \\ &&  \hskip -15mm \nonumber
\left(\begin{array}
{cccccccccc}
     b_1(\lambda )& \mv a_1(\lambda )     &  0      & 0     & 0    & \dots     & 0    & 0    & 0    & 0    \\
     b_1 (\lambda )& \mv a_1  (\lambda )   &  0      & 0     & 0    & \dots    &  0    &  0   & 0    &  0   \\
     0&   0            & \mv b_2(\lambda )   & \mv a_2(\lambda ) & 0    & \dots    &  0   &   0   &  0   &  0   \\
     0&   0            & \mv b_2 (\lambda )  & \mv  a_2 (\lambda )& 0    & \dots    &  0   &  0    &  0   &  0   \\
\multicolumn{10}{c}{\dotfill}                                                                   \\
     0& 0     &  0   &  0   & 0       & \dots     & 0    &  0    &  \mv  b_N(\lambda )  & \mv  a_N(\lambda )     \\
     0& 0     &  0   &  0   & 0       & \dots     & 0    &  0    &  \mv  b_N(\lambda )  & \mv  a_N(\lambda )     \\
     0&  0    & 0    &  0   & 0       & \dots     & 0    &  0    & 0        & 0
\end{array}\right)
\end{eqnarray}
a wavelength dependent $(2N+1,N)$ matrix. It has the elements
\begin{eqnarray}
a_i(\lambda )&=&(1-t_i(\lambda )  +  r_i(\lambda ) )/\tau_i(\lambda )  - t_i(\lambda )
\\
b_i (\lambda ) &=&(t_i (\lambda ) - r_i (\lambda ) -1)/\tau_i (\lambda ) +(1-r_i(\lambda ) ).
\end{eqnarray}
The transfer equation further has
\begin{equation}
{\bf B}=(B(z_1,\lambda ),B(z_2,\lambda ),\dots,B(z_N,\lambda ))^T
\end{equation}
and
\begin{equation}
{\bf I}_{bc}=(I_{sun}(\lambda ),0,0,\dots,0)^T,
\end{equation}
where $I_{sun}$ is the solar irradiation flux.

The vector of the mean intensities is now given by
\begin{equation}
{\bf J}= {\bf \Psi} {\bf I}+{\bf I}_{bc}/2,
\end{equation}
with the wavelength independent $N,(2N+1)$ matrix
\begin{equation}
{\bf \Psi}=\frac{1}{2} \left(\begin{array}
{cccccccccc}
     1& 0      &0   & 0    & 0    & \dots     & 0    & 0    & 0    & 0    \\
     0& 1      & 1  & 0    & 0    & \dots    &  0    &  0   & 0    &  0   \\
     0& 1      & 1  &  0   & 0    & \dots    &  0   &   0   &  0   &  0   \\
     0& 0      & 0  &  1   & 1    & \dots    &  0   &  0    &  0   &  0   \\
\multicolumn{10}{c}{\dotfill}                                                     \\
     0& 0     &  0   &  0   & 0    & \dots     & 1    &  1    & 0   &  0  \\
     0& 0     &  0   &  0   & 0    & \dots     & 0    &  0    & 1   & 1   \\
     0&  0     & 0    &  0   & 0    & \dots    & 0    &  0    & 1   & 1
\end{array}\right),
\end{equation}
such that the energy equation reads in this nomenclature
\begin{eqnarray}
\label{enerdiscont}
\int_{\lambda _\mathrm{cut}}^\infty \sgreekbf{\kappa}(\lambda )\left[{\bf B}(\lambda )-{\bf \Psi}\left({\bf 1}-{\bf M}(\lambda )\right)^{-1}{\bf \Phi}(\lambda ){\bf B}(\lambda )\right]d\lambda 
\nonumber \\
=\int_0^{\lambda _\mathrm{cut}}\sgreekbf{\kappa}(\lambda ){\bf \Psi}\left({\bf 1}-{\bf M}(\lambda )\right)^{-1}{\bf I}_{bc}(\lambda )d\lambda. 
\end{eqnarray}

\subsection*{Boundary conditions}
\noindent
{\bf The lower boundary condition}: We assume that the energy exchange between  surface and the atmosphere is in equilibrium, namely:
\begin{equation}
\sigma \Tsur^4+aI_-(0)+h_\mathrm{sur}(\Tsur-T_\mathrm{air})=I_-(0),
\end{equation}
where $I(z)_-$ and $I(z)_+$ are the downward and upward specific intensities respectively. 
$h_\mathrm{sur} $ is the heat transfer coefficient between the surface and the air through the surface boundary layer. 

\vskip 0.25cm \noindent
{\bf The top boundary condition}: The top boundary condition is the incident stellar radiation, namely
\begin{equation}
I_-(z=z_\mathrm{atm})=I_\mathrm{given, stellar}.
\end{equation}
The outgoing flux is not specified and the converged solution should yield an equilibrium, namely vanishing net energy absorption by the planetary atmosphere. 

\vfill \eject 
\subsection*{Deviations from LTE}

\begin{figure}
\center{\epsfig{file=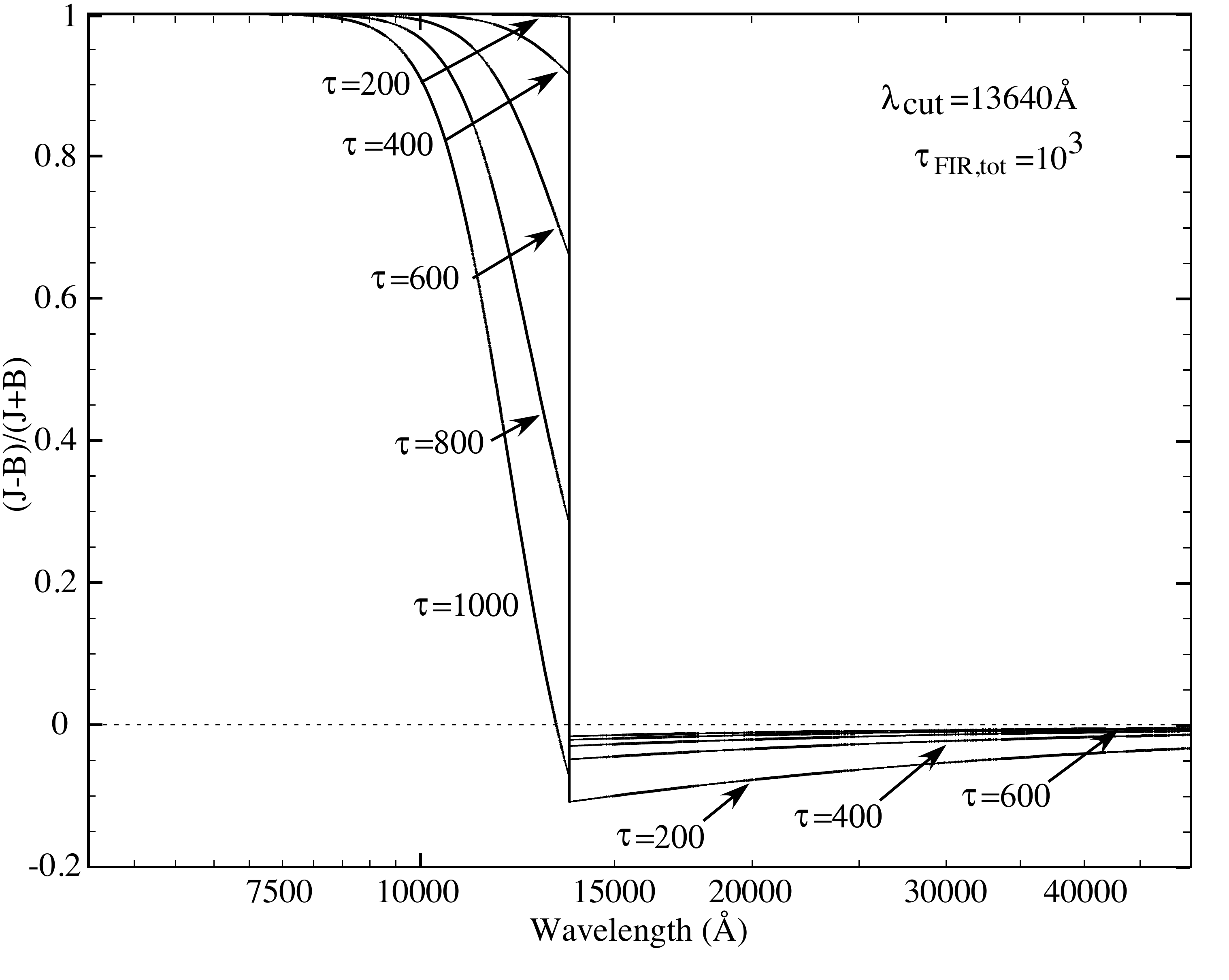,width=2.7in,width=0.45\textwidth  }}
\caption{\figstyle The deviation from LTE for several optical depths in the particular case of $\lambda_{cut}=13640{\rm \AA}$. Despite the very large optical depth, the deviations from LTE are non-negligible. The leakage of the thermal surface radiation into the short-wavelength range varies with height (optical depth), as does the effect of the insolation in the long-wavelength range. }
\label{fig:LTEdeviation}
\end{figure}

The numerical solution allows us to include deviations from LTE as is demonstrated in fig.\ \ref{fig:LTEdeviation}, in the case of $\tau_{\SW} \ll 1$ and $\tau_{\FIR} \gg 1$. Plotted is the function $\psi=(J-B)/(J+B)$ where $J=\oint Id\Omega /4\pi.$  In the visible range $J \gg B$ so that $\psi \approx 1$. If the thermal radiation were in equilibrium, then $\psi=0$. However, we find major deviations close to $\lambda \sim \lambda_\mathrm{cut}$. The surprise is that even at depths of $\tau_{\FIR}\sim 200$ we notice significant deviations.  

Clearly, if there are windows in the IR (or VIS), then similar deviations are expected around the discontinuities in the optical depth.

\end{document}